\documentclass[11pt,leqno,oneside]{extarticle}

\usepackage{cite}

\usepackage{amsmath}
\usepackage{amssymb}
\usepackage{amsfonts}
\usepackage{setspace}
\usepackage{ccfonts} 
\usepackage[varg]{txfonts}
\usepackage{latexsym}
\usepackage{mathrsfs}
\usepackage{epsfig}
\usepackage{psfrag}
\usepackage{color}

\begin{document}

\newcommand{\bm}[1]{\mbox{\boldmath $#1$}}
\def\defi{:=}

\newcommand{\vs}{\vspace{0.2cm}}
\newcommand{\n}{\noindent}
\newcommand{\be}{\begin{equation}}
\newcommand{\ee}{\end{equation}}
\newcommand{\ben}{\begin{displaymath}}
\newcommand{\een}{\end{displaymath}}
\newcommand{\ep}{\hspace{\stretch{1}}$\Box$}
\newcommand{\Da}{{\mathcal{D}}}

\newcommand{\rc}{\color{red}}
\newcommand{\gc}{\color{blue}}
\newcommand{\pc}{\color{magenta}}

\newtheorem{Remark}{Remark}
\newtheorem{Definition}{Definition}
\newtheorem{lemma}{Lemma}
\newtheorem{proof}{Proof}
\newtheorem{Theorem}{Theorem}
\newtheorem{corollary}{Corollary}
\newtheorem{Proposition}{Proposition}
\newtheorem{Corollary}{Corollary}


\newcommand{\Int}{{\rm Int}}
\def\Ec{E^{+}_{0}}
\def\W{{\cal W}}
\def\D{D}
\def\SS{{\cal S}}
\def\gam{{h_S}}
\def\gambar{{\overline{h}_S}}
\def\m{\nu}
\def\IntSigma{{\Sigma^{\circ}}}
\def\IntN{{\N^{\circ}}}
\def\IntM{{\M^{\circ}}}
\def\partialN{\partial^N}
\def\Y{{\mathsf Y}}
\def\nn{{\mathsf n}}
\def\K{{\mathcal K}}
\def\rhoh{\hat{\rho}}
\def\Tau{{\cal T}}
\def\Tauh{\hat{\Tau}}
\def\wh{\hat{w}}
\def\Ione{I_1}
\def\n{\noindent}
\def\Rg{R_g}
\def\Ricbf{\bm Ric}
\def\Ric{Ric}
\def\Ricg{{\Ric_g}}
\def\Rich{{\Ric_h}}
\def\nablah{\nabla}
\def\Deltah{\Delta}
\def\Hess{\mbox{Hess}}
\def\hbar{\overline{h}}
\def\gammabar{\overline{\gamma}}
\def\Richbar{\Ric_{\hbar}}
\def\Ein{\mbox{G}}
\def\Einbf{\mbox{\bf G}}
\def\nablahbar{\overline{\nabla}}
\def\L{{\cal L}}
\def\H{{\cal H}}
\def\tbd{\partial^T}
\def\bd{\partial}
\def\idfull{(\Sigma,g,K;\rho,{\bf J})}
\def\kidfull{(\Sigma,g,K;N,Y;\rho,{\bf J},\Tau)}
\def\id{(\Sigma,g,K)}
\def\kid{( \Sigma,g,K;N,Y)}
\def\SigmaY{\Sigma^{Y}}
\def\tr{\mbox{tr}}
\def\g{\bm{g}}
\def\gD{\bm{g_D}}
\def\gS{\bm{g_S}}
\def\SigmaT{\Sigma^{T}}
\def\B{B(t,S_r)}
\def\Bp{B(t(p),S_r)}

\def\M{{\mathcal{M}}}
\def\N{{\mathcal{N}}}
%
%

\def\Journal#1#2#3#4#5#6{#1, ``#2'', {\it #3} {\bf #4}, #5 (#6).}
\def\JGP{\em J. Geom. Phys.}
\def\JDG{\em J. Diff. Geom.}
\def\CQG{\em Class. Quantum Grav.}
\def\JPA{\em J. Phys. A: Math. Gen.}
\def\PRD{{\em Phys. Rev.} \bm{D}}
\def\GRG{\em Gen. Rel. Grav.}
\def\IJT{\em Int. J. Theor. Phys.}
\def\PR{\em Phys. Rev.}
\def\RMP{\em Rev. Mod. Phys.}
\def\MNRAS{\em Mon. Not. Roy. Astr. Soc.}
\def\JMP{\em J. Math. Phys.}
\def\DG{\em Diff. Geom.}
\def\CMP{\em Commun. Math. Phys.}
\def\APP{\em Acta Phys. Polon.}
\def\PRL{\em Phys. Rev. Lett.}
\def\ARAA{\em Ann. Rev. Astron. Astroph.}
\def\ANP{\em Annals Phys.}
\def\AP{\em Ap. J.}
\def\APJL{\em Ap. J. Lett.}
\def\MPL{\em Mod. Phys. Lett.}
\def\PREP{\em Phys. Rep.}
\def\AASF{\em Ann. Acad. Sci. Fennicae}
\def\ZP{\em Z. Phys.}
\def\PNAS{\em Proc. Natl. Acad. Sci. USA}
\def\PLMS{\em Proc. London Math. Soth.}
\def\AIHP{\em Ann. Inst. H. Poincar\'e}
\def\ANYAS{\em Ann. N. Y. Acad. Sci.}
\def\SPJ{\em Sov. Phys. JETP}
\def\PAWBS{\em Preuss. Akad. Wiss. Berlin, Sitzber.}
\def\PPLL{\em Phys. Lett. A }
\def\QJRAS{\em Q. Jl. R. Astr. Soc.}
\def\CR{\em C.R. Acad. Sci. (Paris)}
\def\CP{\em Cahiers de Physique}
\def\NC{\em Nuovo Cimento}
\def\AM{\em Ann. Math.}
\def\APP{\em Acta Physica Polonica}
\def\BAMS{\em Bulletin Amer. Math. Soc}
\def\CPAM{\em Commun. Pure Appl. Math.}
\def\PJM{\em Pacific J. Math.}
\def\ATMP{\em Adv. Theor. Math. Phys.}
\def\PRSLA{\em Proc. Roy. Soc. London A.}
\def\APPT{\em Ann. Poincar\'e Phys. Theory}
\def\CM{\em Contemp. Math.}
\def\Ast{\em Ast\'erisque}

\begin{center}
{\huge Global and uniqueness properties of stationary 
and static spacetimes with outer trapped 
surfaces}

\vspace{0.5cm}
{\sc\large Marc Mars}

{\small Facultad de Ciencias, Universidad de Salamanca\\
marc@usal.es}\\

\vs
{\sc\large Martin Reiris}\\
{\small Albert Einstein Institut, Max Planck Institut f\"ur Gravitationsphysik\\
martin@aei.mpg.de} \\

\vspace{0.7cm}
\begin{minipage}[c]{13cm}
\begin{spacing}{.7}
{\small Global properties of maximal future Cauchy developments of
stationary, $m$-dimensional asymptotically flat initial data with 
an outer trapped boundary are analyzed. We prove that, whenever
the matter model is well posed and satisfies the null energy condition,
the future Cauchy development of the data is a black hole spacetime.
More specifically,
we  show that the future Killing development
of the exterior of a sufficiently large sphere in the initial data set can be isometrically
embedded in the maximal Cauchy development of the data. In the static setting
we prove, by working directly on the initial data set, that all Killing prehorizons are embedded
whenever the initial data set has an outer trapped boundary and satisfies the null energy condition.
By combining both results we prove a uniqueness theorem for static initial data sets with outer
trapped boundary.}
\end{spacing}
\end{minipage}
\end{center}

\begin{center}
\begin{minipage}[c]{12.5cm}
\tableofcontents
\end{minipage}
\end{center}

\section{Introduction and overview of the main results}

In this paper we investigate the relationship between
asymptotically flat stationary initial data sets with outer
trapped boundary and black holes. By Penrose singularity 
theorem \cite{HE} initial data configurations 
with an outer trapped boundary lead to maximal Cauchy developments
which are geodesically incomplete. On the other hand, physical 
arguments of predictability lead to the weak cosmic censorship  conjecture (see
e.g. \cite{CosmicCensor}) which asserts that any singularity that forms
during a process of gravitational collapse must be a black hole  space-time. 
Black hole space-times (from now on
simply {\it black holes}) satisfy strong
global properties (see below) and it is a very difficult task in general
to determine whether the global properties will be satisfied
knowing only the initial configuration of the spacetime. 
 In the stationary setting, where to a certain sense there is no evolution
at all, the problem must necessarily be much simpler. At first sight one
might even think that, in fact, determining whether the maximal Cauchy 
development of a stationary initial configuration with trapped boundary is a black hole should 
be direct because one would only need to propagate the initial information
by the isometry. The problem is not nearly as simple
because even being able to carry over the development by the isometry
no global information, for example global hyperbolicity, is a priori evident.

There exist at least two different approaches to show that the maximal 
globally hyperbolic development of a stationary initial data
set is a black hole. The first approach tries to prove directly that
the maximal globally hyperbolic development of the data enjoys sufficient
global properties to qualify as a black hole. This is the approach 
we follow in the first part (Section \ref{Stationary}) of this paper.

There exist several definitions of black hole, 
a priori non equivalent, but following the same underlying principle. In a stationary setting, a convenient definition requires the existence
of an asymptotically flat $(m+1)$-end with the property
that  its causal past
does not cover the whole space-time manifold. An asymptotically flat $(m+1)$-end
is the natural generalization to higher dimensions of the usual
notion of asymptotically flat four-end, which can be found e.g.
in \cite{BeigChrusciel}. In essence, the definition demands
that the $(m+1)$-end is topologically the product of a real line times
the exterior of a closed ball in $\mathbb{R}^{m}$. The spacetime metric
is required to be constant along the $\mathbb{R}$ factor and to
approach the Minkowski metric (with time
along the $\mathbb{R}$ factor) at suitable rate.
Since we are interested only in the future of a Cauchy surface,
we restrict the topology of the $(m+1)$-end to be $\mathbb{R}^{+}$
times the exterior of a closed ball in $\mathbb{R}^m$.
Under suitable conditions,
this definition is equivalent to the definition using conformal
compactifications (see e.g. the Appendix of \cite{DamourSchmidt} for the
vacuum, four dimensional
case or Proposition 1.9  in \cite{ChruscielFolclore}
for electrovacuum). 

Concerning stationary initial data and black holes we will prove in Section \ref{Stationary}
\begin{Theorem}
\label{ThmIntro}
Let $\Da$ be an $m$-dimensional ($m \geq 3$)
asymptotically flat stationary initial data with well posed matter model satisfying the 
null energy condition
Suppose that $\partial \Sigma\neq \emptyset$ is future outer trapped. Then the maximal
Cauchy development  $(\M,\g)$ of $\Da$ is a black hole spacetime.
\end{Theorem}
A more precise statement is given in 
Theorem \ref{Thm1} in Section \ref{theresults1},
where we prove  that the Killing development of the data
outside a sufficiently large sphere 
can be isometrically embedded into the
maximal globally hyperbolic spacetime generated by the whole data. 

The second approach to prove that equilibrium initial configurations
lead to black holes is via uniqueness theorems.
In the stationary setting, black hole space-times satisfy uniqueness theorems, in particular for
arbitrary dimension in the static case and in dimension four in the non-static
electrovaccum case under suitable hypotheses \cite{ChruscielCosta}. 
Thus, if one expects that certain stationary initial data develops well behaved black-hole space-times
then such data should be 
embeddable in one of the stationary/static black holes
allowed. In other words the data (inside some region) should be one among those data endowed on sections of
the listed black holes. The hope is that, somehow, such information should be extractable from the initial data
itself to deduce, a fortiori, that the given initial data gives rise to a black-hole space-time.

This strategy has been successfully applied in the past under suitable restrictions.
The first result along these lines is due to P. Miao \cite{Miao2005}
who proved a uniqueness result for asymptotically flat,
three-dimensional vacuum and time-symmetric static Killing initial data
having an outermost minimal boundary. More precisely, Miao 
proved that the data must be isometric to the $\{ t = \mbox{const.}\}$ slice of 
the Schwarzschild spacetime for some mass $M > 0$. A related result was 
found by Carrasco \& Mars 
\cite{CarrascoMars2008, CarrascoMars2011} for data with outer trapped boundary in the case of non-zero second fundamental form
and for more general matter models, provided a number conditions were
satisfied. The generalization to non-vanihing second fundamental form
is relevant because, in the absence
of global information about the spacetime generated by the initial 
data, globally defined time-symmetric slices may simply not exist 
in the spacetime under consideration. Although of interest, the
results in \cite{CarrascoMars2008, CarrascoMars2011} are not fully
satisfactory because they required a number of hypotheses
that were basically dictated by the method of proof,  with no 
fundamental reason to believe that they should be necessary. 
One such hypothesis excluded the presence of so-called non-embedded Killing prehorizons.
A Killing prehorizon is a null immersed
hypersurface where the Killing vector is null, nowhere zero and tangent
(see \cite{ChruscielCosta}). When the surface gravity of the horizon vanishes,
it is a priori possible that the Killing prehorizon is not embedded (see
the discussion in the Addendum of \cite{Cc}). 
In a black hole context (more precisely,
assuming that the domain of outer communications is 
globally hyperbolic) Chru\'sciel and Galloway 
\cite{ChruscielGalloway} have proved that all prehorizons contained in the
domain of outer communications must be embedded. Therefore, in the light of the discussion above, 
one expects to be able to rule out non-embedded prehorizons at the initial data level. 
  
Our second aim in this paper, developed in Section \ref{Static}, is precisely to exclude the existence of 
non-embedded Killing prehorizons
in the exterior region of a static Killing initial data set.
Note that since our global statement in the first part of the paper
is only to the future, we cannot rely 
on the results by Chru\'sciel and Galloway mentioned above. 
This has the advantage that not even the existence
of a spacetime containing the initial data needs to be assumed. 
This means, in particular, that 
no field equations whatsoever are required for this part
of the work. The only requirement we make is that 
the matter model satisfies the null energy condition. 
In this part of the paper, however, we restrict
ourselves to static Killing initial data.
It is an interesting open question whether
the method extends to the stationary (non-static) setting as well. 

Concerning the relation between static Killing initial data and horizons we will be proving in Section \ref{Static}
\begin{Theorem}
\label{Thm2v1}
Let $\Da$ be an $m$-dimensional ($m \geq 3$)
asymptotically flat static Killing initial data set satisfying the null energy condition.
Suppose that the exterior, connected region where the Killing vector is timelike lies in the interior of $\Sigma$, then each degenerate
horizon is a compact embedded submanifold.
\end{Theorem}
Combined with Theorem \ref{ThmIntro}, this
result implies the non-existence of non-embedded Killing prehorizons
in static, asymptotically flat spacetimes with Cauchy surface having an
outer trapped boundary. 

The third, and final, part of the paper, developed in Section \ref{uniq},
is an application of the previous two and establishes a 
uniqueness theorem for static, asymptotically flat initial data with outer trapped
boundary for suitable matter models.  In the vacuum case, the statement is as follows
\begin{Theorem}
\label{uniqueness vacuum}
Let $\Da$ be a static, vacuum $m$-dimensional ($m \geq 3$) asymptotically
flat Killing initial data with non-empty future outer trapped boundary.
Then, the initial data  restricted to the exterior, connected
region where the Killing vector is timelike  can be isometrically embedded in a
Schwarzschild $(m+1)$-dimensional spacetime of mass $M>0$.
\end{Theorem}
The non-vacuum case is treated in Theorem \ref{uniqueness} 
in Section \ref{uniq}. This statement
gives a satisfactory answer to the problem of uniqueness
for static initial data sets with an outer trapped boundary.

\section{Stationary Killing initial data}
\label{Stationary}

\subsection{Background and definitions}
\label{background}
In this paper manifolds are defined to be 
smooth, Hausdorff, connected and paracompact
(hence second countable). Fields on manifolds are assumed to be smooth.
For manifolds with boundary $\Sigma$
we use $\partial \Sigma$ for the boundary and
$\IntSigma (=\Sigma\setminus \partial \Sigma)$ for the
usual notion of interior of a manifold with boundary.
For arbitrary subsets $U$ in a manifold
we use $\overline{U}$ for the topological
closure, $\mbox{Int}(U)$ for the interior and 
$\tbd U$ for the topological boundary. 

We work with $(m+1)$-dimensional spacetimes $(\M,\g)$ ($m \geq 3$).
For a subset $U \subset \M$
we define the {\it null boundary}
$\partialN U$ as the subset of points $p \in \tbd U$ 
such that
there exists a future directed null geodesic segment $\gamma(s)$ of $(\M,\g)$ 
starting at  $p$ and fully contained in $\tbd U$.
The causal past of a set $U$ is denoted by $J^-(U)$ and the future 
domain of dependence of $U$ is denoted by $D^{+} (U)$. 
The conventions we use for these objects, and  for
causality notions in general, follow \cite{Wald}. In particular, 
a spacetime $(\M,\g)$ is globally hyperbolic if it admits
a Cauchy hypersurface $\Sigma$. 
We denote by $\M^+$ the
future domain of dependence of $\IntSigma$.
Note that $\M^{+}$ is a manifold with the smooth boundary $\IntSigma 
$. Let $n$ be the space-time future directed unit
normal to $\IntSigma$. The induced metric on $\IntSigma$
will be denoted by $g$ and the second fundamental form by $K$ (in the direction of $n$). If $\Sigma$ has boundary we assume it is compact 
and that both $g$ and $K$ extend smoothly to $\partial \Sigma$. 

Let $\Einbf$ be the Einstein tensor of $\g$, namely 
$\Einbf \defi  \Ricbf - \frac{1}{2} {\bm R} \g$,
where $\Ricbf$ denotes the Ricci tensor of the metric $\g$ and ${\bm R}$ is the curvature scalar
(our sign conventions are such that the Ricci tensor and curvature scalar of a round sphere are positive) 

We define $\rho$, $J$ (1-form) and $\Tau$ (symmetric two-tensor) at $p\in \Sigma$ according to 
\begin{align}
& \rho= \Einbf (n,n), \nonumber \\
& J(v)= - \Einbf (n,v),\ v\in T_{p}\Sigma, \label{Projection} \\
& \Tau(v,w)= \Einbf (v,w),\ v,w\in T_{p}\Sigma. \nonumber 
\end{align}

\begin{Remark}
$\rho$, $J, \Tau$ are defined in terms of the Einstein tensor,
and not as components of any energy-momentum tensor because we will not assume any specific field equations relating
the Einstein tensor with the energy-momentum tensor of the matter fields. Neither the matter model
nor the field equations will be of concern to us except
that we will require a {\it well posedness property} defined later.
\end{Remark}

The data $(\Sigma;(g,K);(\rho,J))$ satisfies the {\it energy} and {\it momentum constraint equations}
\begin{align}
&\label{C1} R_{g}-|K|_g^{2}+k^{2}=2\rho,\\
& \label{C2} {\rm div}_g \ (K - k g)=-J.
\end{align}  
where $k={\rm tr}_{g} K$ and ${\rm div}_{g}$
is the divergence with respect to  the metric $g$.

The boundary of $\Sigma$ (if any) is said to be {\it future outer trapped} if it is
compact and 
\begin{equation*}
\theta^+ (\partial \Sigma) \defi \tr\big |_{\partial \Sigma} K + H<0.
\end{equation*}

\n where ${\rm tr}\big|_{\partial \Sigma} K$ is the trace of $K$
restricted to $\partial \Sigma$ and $H$ is the mean curvature of $\partial
 \Sigma$ in the inward
direction to $\Sigma$.

We will assume that we have a Killing vector field $\xi$ on $(\M,\g)$ having the decomposition $\xi= Nn+Y$  along $\Sigma$. $(N,Y)$ will be part of the initial data. Because $\xi$ is Killing we have, over $\Sigma$, the equations (see
e.g. \cite{Maerten})
\begin{align}
{\mathcal{L}}_{Y} g  & = -2NK,\label{kid1}\\
{\mathcal{L}}_{Y} K  & =-  \Hess_g N + N
\left ( \Ricg + k \, K-2K\circ K \right ) - N \left 
(\Tau-\frac{1}{m-1}({\rm tr}_{g}\Tau-\rho)  \right ) g\label{kid2}. 
\end{align}
where ${\mathcal L}$ denotes Lie derivative,
$\Hess_g$ is the Hessian with the metric $g$,
$\Ricg$  is the Ricci tensor of $g$ and $K \circ K$
is the tensor obtained from $K \otimes K$ by tracing with $g$ the second
and fourth indices. Concerning the data we make the following definition.

\begin{Definition}{\rm (AF-KID)}
\label{AF-KID}
The data $\Da \defi
(\Sigma;(g,K);(\rho,J,\Tau);(N,Y))$ is said to be a 
\begin{enumerate}
\item  Killing initial data (KID) if it satisfies (\ref{C1})-(\ref{C2}) and (\ref{kid1})-(\ref{kid2}),
\item and stationary asymptotically flat (AF) if there is a compact set 
whose complement $\Sigma^{\infty}$
is diffeomorphic to $\mathbb{R}^{m+1} \setminus \{ \mbox{closed ball } \}$
and, in the Euclidean coordinates $\bar{x}
=(x^{1},\cdots,x^{m+1})$ on $\Sigma^{\infty}$
defined by the diffeomorphism we have
\begin{align}
 &  g_{ij}-\delta_{ij} =O^{2} \left ( 1/r^{m-2} \right ), \quad \quad  & K_{ij}=O^{2} \left ( 1/r^{m-1} \right ),
\\
 &  N-N_{\infty}=O^{2} \left ( 1/r^{m-2} \right ), \quad \quad &  Y_i - Y_{\infty \, i}=O^{2} \left ( 1/r^{m-2} \right ) . \label{decay}
\end{align}
and with $r =  |\bar{x} | \defi \sqrt{(x^1)^2 + \cdots (x^m)^2}$,
where $N_{\infty}$, $Y_{\infty}$ are constants satisfying 
$N_{\infty} > | Y_{\infty} |$.
\end{enumerate}

\n Under these conditions we will say simply that $\Da$ is a stationary asymptotically flat initial data.  

\end{Definition}

Let $\lambda \defi N^{2}-|Y|_g^{2}$. Since $\lambda \rightarrow N^2_{\infty} 
- | Y_{\infty} |^2 >0$ at infinity, we can assume
(after restricting $\Sigma^{\infty}$
if necessary) that $\lambda > 0$ on the asymptotically flat end
$\Sigma^{\infty}$. 
We denote by $\Sigma^{T}$ ($T$ from time-like) the connected component of $\{\lambda>0\} \subset \Sigma$ containing $\Sigma^{\infty}$. 

The {\it Killing development} of subregions of $\Sigma^{T}$ is defined as follows.

\begin{Definition}{\rm (Killing developments)}
\label{KillDev} 
Let $\Omega$ be a connected open subset of $\Sigma^{T}$. Then the infinite Killing development of the data at 
$\Omega$ is defined as the space-time
\begin{equation}
\K(\Omega)\defi \left (\Omega\times (0,\infty) 
,\gD=-\lambda dt^{2}+ \Y \otimes dt + dt \otimes \Y +g \right ), \label{development}
\end{equation}
where $\Y \defi  g (Y,\cdot \, )$. The restriction of $\K(\Omega)$ to
$t\in (0,\bar{t}\,]$ ($\bar{t}>0$) is denoted by $\K(\Omega,\bar{t})$.

\end{Definition}

The Killing development is a space-time with Killing field $\xi=\partial_{t}$ and Einstein tensor
$\Einbf$  as defined above from $\rho,J$ and $\Tau$. 

In this paper we will also use a related notion of Killing development
of hypersurfaces $V$ (with or without boundary) embedded 
in a spacetime $(\M,\g)$ admitting a Killing vector $\xi$. The 
only requirement is that $V$ is everywhere transverse to $\xi$. 
If we denote by $g$ the first fundamental form of $V$, by
$\Y$ the pull-back of the one-form obtained by lowering the index
to $\xi$ and $\lambda \defi - \langle \xi, \xi \rangle $ (where
$\langle \, , \, \rangle $ denotes scalar product with the spacetime
metric $\g$) then, the Killing development $\K (V)$ is defined
as the spacetime $V \times (0,\infty)$ with the metric
$\gD$ defined exactly as in (\ref{development}). It is immediate to
see that, $\xi$ being transverse to $V$ everywhere, $\gD$ is a metric
of Lorentzian signature. There is no restriction on the causal
character of $V$, which in particular is allowed to be null. Note that
when  the Killing initial data is embedded in a spacetime, we have a priori two 
ways of contructing the Killing development. However, the spacetimes
constructed with either definition are the same, so no confusion
arises. We also emphasize  that the Killing development
$\K (V)$ is an abstract spacetime defined on its own which, a priori,
has nothing to do with the original spacetime $(\M,\g)$.

Given $(\M,\g)$ with a Killing vector $\xi$, let
$\beta_q(\lambda)$, $\lambda\geq 0$, be the Killing orbit starting at $q \in \M$ (i.e.
$\beta_q (\lambda=0)=q$).
For any $W\subset \M$ and $0 \leq a \leq b$ satisfying the property
that all the Killing orbits $\beta_q (\lambda), q \in W$
extend to all 
values $\lambda \in [a,b]$ we  will denote by $O_{[a,b]}(W)$
the set
\begin{equation*}
O_{[a,b]}(W) \defi \{ \beta_q (\lambda)\in \M  / q \in W, a \leq \lambda
\leq b \}.
\end{equation*}
If $a=b$ we will simply write $O_{[a]}(W)$ for $O_{[a,a]} (W)$.
In the following it will be convenient to use
two different notations for
Killing orbits: $\beta_{q}(\lambda)$ when the orbit
satisfies $\beta_q(\lambda=0) =q$ and 
$\lambda$ takes positive values and $\alpha_p (\mu)$
when the orbit satisfies $\alpha_{p} (\mu=0) = p$ and
$\mu$ takes negative values.

\begin{Definition} {\rm (Well posed matter models)} Let $\Da$
be a Killing initial data. We say that the matter model is well posed if the field
equations are such that a Killing initial data generates a unique maximal globally hyperbolic
space-time $(\M,\g)$ having a Killing field $\xi$ extending $\xi\big |_{\Sigma}=Nn+Y$. 
\end{Definition}

As usual, we will refer to the maximal globally hyperbolic spacetime $(\M,\g)$  as
{\it maximal Cauchy development}. 
The simplest example of well posed matter model is vacuum, i.e. when
the field equations are $\Einbf =0$. In this case, the 
Cauchy problem is well posed and the maximal Cauchy development
admits a Killing vector \cite{Moncrief}. The same is true e.g.
in electrovacuum 
\cite{ChruscielFolclore} 
 and
for many other matter models \cite{Rendall}.

We will say that a matter model satisfies the null energy
condition if all Cauchy developments $(\M,\g)$
solving the field equations satisfy 
$\Einbf (l,l)\geq 0$ for any null vector $l$.

We need some observations on isometric embeddings. Let 
$(\N_{i},\g_{\N_{i}})$, $i=1,2$ be two connected Lorentzian or Riemannian manifolds (possibly with smooth boundary). 
Let ${\cal W}_{i}\subset \N^{\circ}_{i}$, $i=1,2$ be two sets such that ${\cal W}_{i}\subset \overline{\mbox{Int}({\cal W}_{i}})$, $i=1,2$ and 
$\Int({\cal W}_{i})$, $i=1,2$ are connected and non-empty. Let 
$p_{i} \in \mbox{Int}({\cal W}_{i})$, $i=1,2$. Let $\phi:T_{p}\N_{1}\rightarrow T_{p'}\N_{2}$ be a
linear isometry between the tangent space to $\N_{1}$ at $p_{1}$ and the
tangent space to $\N_{2}$ at $p_{2}= \phi(p_{1})$. Then we will write $(\N_{1},{\cal W}_{1})\Subset_{\phi} (\N_{2},{\cal W}_{2})$ if there is an open set $U_{1}$ of $\N_{1}$ containing ${\cal W}_{1}$ and a smooth map 
$\varphi:U_{1}\rightarrow \N_{2}$ such that $\varphi\big|_{\mbox{Int}{\cal W}_{1}}$ is an isometric embedding
of ${\mbox{Int}({\cal W}_{1})}$ into ${\mbox{Int}({\cal W}_{2}}$).
In such case $\varphi\big|_{{\cal W}_{1}}$ is unique (and determined only by $\phi$). In the rest of the work we will simply write ${\cal W}_{1}\Subset {\cal W}_{2}$ and it should be understood that this also entails the existence of companion manifolds $\N_{1}$, $\N_{2}$ and map $\phi$. The companion objects will be understood from the context. Note the transitivity property: ${\cal W}_{1}\Subset {\cal W}_{2}$, ${\cal W}_{2}\Subset {\cal W}_{3}\Rightarrow\ $ ${\cal W}_{1}\Subset {\cal W}_{3}$. When ${\cal W}_{1}\Subset {\cal W}_{2}$ we will say that ${\cal W}_{1}$ {\it lies} in ${\cal W}_{2}$.

The uniqueness of the maximal future globally hyperbolic
space-time $\M^{+}$ implies that any future globally
hyperbolic space-time $(\N,\g_{\N})$ with Cauchy hypersurface
$C =\partial \N$ isometric to a particular connected open subset 
of $(\Sigma,g)$, lies uniquely inside
$(\M^{+},\g$). In other words $\N\Subset
\M^{+}$. This fact will be of fundamental importance.


\vs We  will denote by $S_r$ the coordinate sphere of coordinate radius $r$ in the asymptotically flat end
$\Sigma^{\infty}$. $S_r$ separates $\Sigma$ into two closed parts (i.e. including
their boundaries), $\Sigma^{E}(r)$ (``E" from ``Exterior") and $\Sigma^{I}(r)$ (``I" from ``Interior").  
Note that $S_r \subset \Sigma^T$ and $\Sigma^E(r) \subset \Sigma^T$.
For $d \geq 0$, we also define $T(\partial \Sigma,d) \defi \{ p \in \Sigma / \mbox{dist}_{g} (p,\partial \Sigma)
\leq d \}$ and $\Sigma_d \defi \{ p \in \Sigma / \mbox{dist}_{g} (p,
\partial \Sigma) \geq d \}$.

\begin{Definition} Let $(\M,\g)$ be a globally hyperbolic space-time with Cauchy surface $\Sigma$ and asymptotically flat stationary data $\Da$. Suppose
$\K(\Sigma^{E}(r))\Subset \M^{+}$. Then we define the future event horizon (over $\Sigma$) as the topological boundary of $J^{-}(\K(\Sigma^{E}(r)))\cap \Sigma$ 
(as a subset of $\Sigma$).
\end{Definition}

The Killing development $\K (\Sigma^E(r))$ is a  future
asymptotically flat $(m+1)$-end as described in the
Introduction. 

\subsection{The statements of the main results: Theorems \ref{Thm1} and \ref{ThmExterior}}
\label{theresults1}

The following theorem is a precise version of Theorem \ref{ThmIntro} in the
Introduction. 
\begin{Theorem}
\label{Thm1}
Let $\Da$ be an asymptotically flat, stationary initial data set with well posed matter model satisfying the null energy condition. Suppose that $\partial \Sigma$ (if non-empty)
is future outer trapped. Then the maximal Cauchy development $(\M,\g)$  of $\Da$ satisfies
\begin{enumerate}
\item
There is $r > 0$ such that
$\K(\Sigma^{E}(r))$ lies in $\M^{+}$. \label{E1T1}
\item There is $d>0$, such that $T(\partial \Sigma, d) 
\cap J^{-}(\K(\Sigma^{E}(r)))=\emptyset$. \label{E2T1} 
\end{enumerate}
\end{Theorem}

\n In basic terms, if the boundary of $\Sigma$ is future
outer trapped, then 
the future event horizon over $\Sigma$ exists,
is computable or constructible from the data and does not intersect
 $\partial \Sigma$. Therefore $\partial \Sigma$ lies inside the
future black hole region.  

The constant $r$ in item 1 is introduced in Proposition \ref{P0} later.
The constant $d$ is any positive  constant satisfying the property that 
$\partial \Sigma_{d'}$, $0\leq d' \leq 2d$, is diffeomorphic
to $\partial \Sigma$ and $\theta^+ (\partial \Sigma_{d'}) < 0$. 

\vs 
We also have
\begin{Theorem}\label{ThmExterior}
Let $\Da$ be an asymptotically flat stationary initial
data with well posed matter model satisfying the null energy
condition.
Suppose that $\partial \Sigma \neq \emptyset $ is future outer trapped.
Then, the exterior of the event horizon contains $\Sigma^{T}$.
In particular, $\Sigma^T \cap \partial \Sigma = \emptyset$. 
\end{Theorem}
\subsection{Structure and proof of Theorem \ref{Thm1}.}\label{OU}

The proof is constructive. We define first the future globally hyperbolic region $\Ec$. The definition will come out from the following proposition which we leave without proof.
\begin{Proposition}\label{P0}
Consider the domains of dependence $D^{+}(\Sigma^{E}(r))$ depending on $r$, as sets inside the Killing development $\K(\Sigma^{E}(r))$.
Then, if $r$ is big enough the null boundary $\partialN D^{+}(\Sigma^{E}(r))$ is a smooth null hypersurface foliated by future complete null geodesic rays starting at $S_r$. Moreover $D^{+}(\Sigma^{E}(r))\Subset \M^{+}$. \end{Proposition}
Fixed one such $r$ we will denote from now on $\Ec \defi D^{+}(\Sigma^{E}(r))$ and its Cauchy surface $\Omega_{0}=\Sigma^{E}(r)$. 
Now, starting from $\Ec$, we will construct inductively a sequence
$\Ec \Subset E^{+}_{1}\Subset E^{+}_{2}\Subset\ldots \Subset E^{+}_{\infty}=\cup_{i\geq 0} E^{+}_{i}$, such that $E^+_i \setminus \partialN E^+_i$  are globally hyperbolic spacetimes with
Cauchy surfaces $\Omega_{0}\Subset \Omega_{1}\Subset \Omega_{2}\Subset \ldots \Subset\Omega_{\infty}\defi \cup_{i\geq 0} \Omega_{i} \Subset \Sigma^{\circ}$ none  touching the tubular neighborhood $T(\partial \Sigma,d)$ for some $d >0$,
namely having $\Omega_{\infty}\cap T(\partial \Sigma,d)=\emptyset$. From the uniqueness of the maximally globally hyperbolic development $\M^{+}$ we conclude that $E^{+}_{i}\Subset \M^{+}$ for all $i$ and therefore that $E^{+}_{\infty}\Subset \M^{+}$. Moreover we show explicitly along the construction that  $\K(\Sigma^{E}(r))\Subset E^{+}_{\infty}$. This, together with $E^{+}_{\infty}\Subset \M^{+}$, gives $\K(\Sigma^{E}(r))\Subset \M^{+}$ which is the claim (\ref{E1T1}) in Theorem \ref{Thm1}. On the other hand $J^{-}(\K(\Sigma^{E}(r)))\cap T(\partial \Sigma,d)\subset E^{+}_{\infty}\cap T(\partial \Sigma,d)=\Omega_{\infty}\cap T(\partial \Sigma,d)=\emptyset$ which is the claim (\ref{E2T1}), and last,  in Theorem \ref{Thm1}. We present now progressively the main definitions and Propositions (auxiliary Propositions \ref{P1}-\ref{P4}) leading to the construction of the sequence $\{E_{i}\}$. The construction, together with the proof of Theorem \ref{Thm1}, is explained after the statement of Proposition \ref{P4}  which is the main statement of this section and which in itself structures the inductive procedure.   

An important collection of regions for the proof are $\W_t$, $t\geq 0$
defined as 
\begin{equation}
\W_t : = \K(\partialN E^+_0,t) \cup E^+_0 \subset \left (\Sigma^E(r) \times
[0,\infty) , \gD \right ).
\end{equation}
Note that $\W_t$ is a Lorentzian manifold with smooth boundary and
corners. Suppose that $\W_t\Subset \M^{+}$, $t\geq 0$. We consider the sets 
\begin{align*}
& E_{t}  \defi J^{-}(\W_t),\\
& E^{+}_{t}  \defi E_{t}\cap \M^{+},
\end{align*}
where here and in the following $J^{-}$ is taken in the spacetime $(\M,\g)$.
When $t=0$ then $E^{+}_{t}=\Ec$ and $\partialN \Ec$ is, as we said, smooth. Moreover we will prove
\begin{Proposition}\label{P1} $\xi$ points strictly outwards from $\Ec$ at $\partialN \Ec$.
\end{Proposition}
Crucially, this property is generalizable to the sets $E_{t}$, $t\geq 0$ which could fail to have smooth boundaries (although their boundaries are Lipschitz manifolds). We prove that $\xi$ {\it points strictly outwards} from $E_{t}$ at $\partialN E_{t}$ in the sense of the following definition which generalizes the standard sense that we understood in Proposition \ref{P1}.
\begin{Definition}\label{D1}\label{SO}
  Suppose that $\W_t\Subset \M^{+}$. Let $E_{t}=J^{-}(\W_t)$. Then, we say that $\xi$ points strictly outwards to $E_{t}$ at $\partialN E_{t}$, if for every $p\in \partialN E_{t}$, there is $\mu_{p}<0$ such that 
\begin{enumerate}
\item[{\rm\small\bf D1-1.}] $\alpha_{p}(\mu)\in \Int (E_{t})$, for all $\mu$ with $\mu_{p}\leq \mu<0$, where $\alpha_{p}(\mu)$ is the Killing orbit passing through $p$ at $\mu=0$,
\item[{\rm\small\bf D1-2.}] there is an $m$-manifold $V_{p}\subset \Int(E_{t})$, transversal to $\xi$ at $\alpha_{p}(\mu_{p})$ such that, if we
denote by $\beta_{q}(\lambda)$, $\lambda\geq 0$ the Killing orbit passing through $q\in V_{p}$ at $\lambda=0$ then 
\begin{enumerate}
\item[{\rm\small\bf D1-2-(a).}] there is a first $\lambda>0$ , denoted by $\lambda_{q}$, for which $\beta_{q}(\lambda_{q})\in \partialN E_{t}$, 
\item[{\rm\small\bf D1-2-(b).}] the map $q\rightarrow \beta_{q}(\lambda_{q})$  from $V_{p}$ into $\partialN E_{t}$ is continuous
\item[{\rm\small\bf D1-2-(c).}] for every $q\in V_{p}$, $\beta_{q}(\lambda)\in (\M\setminus E_{t})$ if $\lambda>\lambda_{q}$ but close to it,
\end{enumerate}
\end{enumerate}
\end{Definition}
In other words $\xi$ points strictly outwards to $E_{t}$ at its boundary if every Killing orbit starting at $\Int(E_{t})$ either remains inside $\Int(E_{t})$ or crosses $\partialN E_{t}$. An obvious consequence of the Definition
is that if an orbit starts in $\Int(E_t)$ and crosses  $\partialN E_t$, then it never returns to $E_{t}$. 

Recalling, we will prove
\begin{Proposition}\label{P2}
Suppose that $\W_t\Subset \M^{+}$. Let $E_{t}=J^{-}(\W_t)$. Then $\xi$ points strictly outwards from $E_{t}$ at $\partialN E_{t}$ in the sense of Definition \ref{SO}. 
\end{Proposition}
On the other hand if $\xi$ points strictly outwards from $E_{t}$ at $\partialN E_{t}$ in the sense of Definition \ref{SO}, then we prove that one can abstractly extend $E_{t}$ ``along the Killing" any time $\bar{t}>0$ in the sense of the following definition which generalizes the notion of Killing development introduced in Definition \ref{KillDev} before.
\begin{Definition}\label{D2} Suppose that $\W_t\Subset \M^{+}$. Let $E_{t}=J^{-}(\W_t)$. The {\it infinite abstract Killing 
development of $E_{t}$}, $K(E_{t})$, is defined as the manifold formed by the open sets $\{\Int(E_{t}), \K({V}_{p}), p\in \partialN E_{t} \}$ endowed with their respective
metrics and subject to the following identifications

\begin{enumerate}
\item[{\rm\small\bf D2-1.}] the points $x_{1}=(q,\lambda)\in 
\K(V_{p}) = V_{p}\times (0,\infty)$ and $x_{2}\in \Int(E_{t})$ are identified iff $\lambda<\lambda_{q}$ and $\beta_{q}(\lambda)=x_{2}$.

\item[{\rm\small\bf D2-2.}] the points $x_{1}=(q_{1},\lambda_{1})\in \K({V}_{p_{1}})$
and $x_{2}=(q_{2},\lambda_{2})\in \K(V_{p_{2}})$ are identified
iff 

\begin{enumerate}
\item[{\rm\small\bf D2-2-(a).}] $\beta_{q_{1}}(\lambda_{q_{1}})=\beta_{q_{2}}(\lambda_{q_{2}})$ and,

\item[{\rm\small\bf D2-2-(b).}] $\lambda_{q_1} - \lambda_1= \lambda_{q_2} - \lambda_2$.
\end{enumerate}
\end{enumerate}
The abstract Killing development at time $\bar{t}\geq 0$ is defined as the following region of $K(E_{t})$
\ben
K(E_{t},\bar{t}) \defi \Int(E_{t})\cup \{\beta_{q}(\lambda+\lambda_{q}), q\in V_{p},p\in
\partialN E_{t},0\leq \lambda \leq \bar{t}\}.
\een
Similarly one defines
\ben
K(E^{+}_{t},\bar{t}) \defi (\Int(E^{+}_{t})\cap \M^{+}) \cup \{\beta_{q}(\lambda+\lambda_{q}), q\in V_{p},p\in
\partialN E^{+}_{t},0\leq \lambda \leq \bar{t}\}. 
\een
\end{Definition}
Of course we have $E^{+}_{t}\Subset E_{t}\Subset K(E_{t},\bar{t})$. However one must think $K(E_{t},\bar{t})$ as a new spacetime bearing a priori no global relation with $\M$. In general one would not expect that $K(E_{t},\bar{t})\Subset \M$. As we will explain in Proposition \ref{P4} the situation will be different for $K(E^{+}_{t},\bar{t})$ if we select $\bar{t}$ properly and this is what will allow us eventually to construct the sequence $\{E^{+}_{i}\}$. 

Recalling, we will prove
\begin{Proposition}\label{P3} Suppose that $\W_t\Subset \M^{+}$. Let $E_{t}=J^{-}(\W_t)$
. Then, for any $\bar{t}\geq 0$ the abstract Killing development $K(E_{t},\bar{t})$ is a smooth, Lorentzian and second countable manifold with null and Lipschitz boundary. 
\end{Proposition} 
\n The following proposition will structure the construction of the sequence $ \{E^{+}_{i}\}$ that we explain thereafter.
\begin{Proposition}\label{P4} There is $t_{*}>0$
depending only on the initial data over $\Sigma_{d/2}\cap \Sigma^{I}(r+1)$ such that if for some $t\geq 0$, we have
\begin{enumerate}
\item[{\rm\small\bf H1.}] 
$\W_t\Subset \M^{+}$,
\item[{\rm\small\bf H2.}]
$T(\partial \Sigma,2d)\cap E^{+}_{t}=\emptyset$ with $E^{+}_{t}\defi J^{-}(\W_t)\cap \M^{+}$,
\end{enumerate}
then,
\begin{enumerate}
\item[{\rm\small\bf C1.}]
$K(E^{+}_{t},t_{*})\Subset \M$,  and therefore $\W_{t+t_{*}}\Subset \M^{+}$.
\item[{\rm\small\bf C2.}] $T(\partial \Sigma,2d)\cap J^{-}(K(E^{+}_{t},t_{*}))=\emptyset$.
\item[{\rm\small\bf C3.}] $E^{+}_{t+t_{*}} \defi 
J^{-}(\W_{t+t_{*}})\cap \M^{+} = J^{-}(K(E^{+}_{t},t_{*}))\cap \M^{+}$
\end{enumerate}
\end{Proposition}
\n We are ready to prove Theorem \ref{Thm1}. 

\vs
\n {\it Proof of Theorem \ref{Thm1}}:  To construct the sequence $\{E^{+}_{i}\}$ we proceed as follows. First, when $i=0$, $\Ec$ is as we defined it before. Now, {\small\bf H1} and {\small\bf H2} hold in Proposition \ref{P4} with $t=0$. Then, conclusion {\small\bf C1} gives $\W_{t_{*}}\Subset \M^{+}$ which is {\small\bf H1} with $t=t_{*}$. In addition conclusions {\small\bf C2} and {\small\bf C3} give $E^{+}_{t_{*}} \cap T(\Sigma,2d)=\emptyset$ with $E^{+}_{t_{*}}=J^{-}(\W_{t_{*}})\cap \M^{+}$ which is {\small\bf H2} with $t=t_{*}$. Then define $E^{+}_{1}= J^{-} (\W_{t_{\star}}) \cap \M^{+}$. Applying repeatedly Proposition \ref{P4}  in this way, we are led to define $E^{+}_{i}$ as $E^{+}_{i}= J^{-} (\W_{i t_{\star}}) \cap \M^{+}$ which is the desired sequence.  \ep

\subsection{Proofs of the auxiliary Propositions \ref{P1}, \ref{P2}, \ref{P3}, \ref{P4} and of the Theorem \ref{ThmExterior}.}
\n {\it Proof of Proposition \ref{P1}}: The Proposition is direct from the fact that, as a field inside $\K(\Sigma^{E}(r))$, $\xi$ is time-like and future-pointing and that $\partialN \Ec$ is smooth and null. An alternative argument (to be used later) comes from the observation that on a smooth null boundary like $\partialN \Ec$, proving that $\xi$ points strictly outwards is equivalent to prove that for any null geodesic $\gamma(\tau)$ in $\partialN \Ec$ (parametrized by affine parameter $\tau$ into the future direction) we have
$\langle \gamma',\xi \rangle <0$. As $\xi$ is a Killing field we have
\ben
\langle \gamma'(\tau),\xi(\gamma(\tau)) \rangle =\langle \gamma'(0),\xi(\gamma(0)) \rangle,
\een
where $\gamma(0)$ is the initial point of the geodesic at $S_r$. As $\xi$ is timelike on $S_r$ it is $\langle \gamma'(0),\xi(\gamma(0)) \rangle <0$. The statement follows.\ep

The following lemma is useful for the proof of Proposition \ref{P2}.

\begin{lemma}
\label{LemP3}
Assume that $\W_t  \Subset \M^+$ and let $E_{t}=J^{-}(\W_t)$. Then, the closure of 
$\Int (E_{t})\cap \partialN \W_t$ is compact. Moreover, any null
geodesic $\gamma$ in $\partialN E_{t}$ satisfies $\langle \gamma',\xi \rangle <0$.     
\end{lemma}

\n {\it Proof}:  Assume that
$\Int (E_{t})\cap \partialN \W_t$ is not compact. 
Then there is a divergent sequence of points $\{q_{i}\}$, $q_{i}\in \partialN \W_t$, a sequence of points $\{p_{i}\}\in \W_t$, and a sequence of timelike and past directed curves $\Gamma_{i}(\tau)$, $\tau\in [0,1]$ (in $\M^{+}$), such that, for every $i\geq 0$, $\Gamma_{i}$ starts at $p_{i}$ and ends at $q_{i}$. However, as $\Gamma_{i}$ is timelike and past directed, we claim that to reach $q_{i}$, $\Gamma_{i}$ must first cross the set $U=\{\beta_{q}(\lambda),q\in \partial \Omega_{0}, 0\leq \lambda\leq t\}$. Indeed to reach $q_{i}$ from its future the curve $\Gamma_{i}$ must first leave $\W_t$, but being timelike and past directed it cannot cross $\partialN\W_t$, nor it can enter $\Ec$ for it could not leave $\Ec$ again. The claim follows. Denote by ${\rm Vol}(\Gamma_{i})(\tau)={\rm Volume}_{\bf g}(J^{-}(\Gamma_{i}(\tau))\cap \M^{+})$. Then ${\rm Vol}(\Gamma_{i})(\tau)$ is, for every $i$, a monotonically decreasing function of $\tau$. Moreover as $\{q_{i}\}$ is a divergent sequence we must have ${\rm Volume}_{\bf g}(J^{-}(q_{i})\cap \M^{+})={\rm Vol}(\Gamma_{i})(1)\rightarrow \infty$. On the other hand as, for every $i$, $\Gamma_{i}$ crosses $U$, and ${\rm Vol}(\Gamma_{i})(\tau)$ is monotonically decreasing, it must be ${\rm Vol}(J^{-}(\Gamma_{i}))(1)\leq \sup \{{\rm Vol}(J^{-}(q)\cap \M^{+}),q\in U\}<\infty$ for all $i\geq 0$, which gives a contradiction. An important consequence of this, to be used later, is that every inextensible future null geodesic
$\gamma$ in $\partialN E_{t}$ 
becomes eventually a null geodesic of $\partialN \W_t$. Constancy of 
$\langle \gamma', \xi \rangle$ along this geodesic proves the claim $\langle \gamma',\xi \rangle <0$.   \ep

\vspace{0.6cm}
\n {\it Proof of Proposition \ref{P2}}: We show first {\small\bf D1-2-(a)}
. Through 
every point $p$ in $\partialN
E_{t}$ there passes a future inextensible null geodesic $\gamma_{p}(\tau)$, $\tau>0$, starting at $p$
and fully contained in $\partialN E_{t}$ (see \cite{Wald}). By Lemma \ref{LemP3}  $\langle \gamma',\xi \rangle < 0$. 
Moreover, every point $p'=\gamma_{p}(\tau)$, $\tau>0$, is a smooth point
of $\partialN E_{t}$. But for smooth points we know that if $\langle 
\gamma'_{p}(\tau),\xi(p') \rangle <0$  then $\xi$ points strictly outwards
to $E_{t}$ at $p'$. If $\tau>0$ is small enough, then $p\in \partial J^{-}(\gamma_{p}(\tau))$ and $p$ is a smooth
point of $\partialN J^{-}(\gamma_{p}(\tau))$. But because $\langle \gamma_{p}'(0),\xi(p) \rangle = \langle \gamma_{p}'(\tau),\xi(\gamma_{p}(\tau))
\rangle <0$, we deduce that $\xi(p)$ points strictly outwards to $J^{-}(\gamma_{p}(\tau))\subset E_{t}$ at $p$.
Since $\Int(J^{-} (\gamma_{p}(\tau)) \subset \Int (E_t)$ there is $\mu_{p}<0$ such that $\alpha_{p}(\mu)\in \Int (E_{t})$ for
$\mu\in [\mu_{p},0)$ thus showing {\small\bf D1-1}. 

\vs
We prove now {\small\bf D1-2-(a)}
. First note that there is $a>0$ and a closed smooth three-submanifold $V_{p}$ (with smooth boundary) transversal to $\xi$  and
containing $\alpha_p (\mu_p)$
such that for every $q\in V_{p}$ the Killing orbit $\beta_{q}(\lambda)$, passing through $q$ at $\lambda=0$, 
extends to all values $\lambda\in [0,-\mu_{p}+a]$. For every $q\in V_{p}$ define $\bar{\lambda}_{q}=\inf\{a,\lambda^{T}_{q}\}$ where $\lambda^{T}_{q}$ is the first $\lambda>0$ such that $\beta_{q}(\lambda)\in \partialN E_{t}$ (we take $\lambda^T = \infty$ if 
$\beta_q(\lambda)$ never intersects $\partialN E_{t}$).
Note that if $\bar{\lambda}_{q}<a$ then $\beta_{q}(\bar{\lambda_{q}})\in \partialN E_{t}$. From this and because $\bar{\lambda}_{\alpha_{p}(\mu_{p})}=-\mu_{p}$ we deduce that if $q\rightarrow \bar{\lambda}_{q}$ is continuous at $q=\alpha_{p}(\mu_{p})$ then one can take a smaller $V_{p}$ if necessary in such a way that $\bar{\lambda}_{q}<a$ for every $q\in V_{p}$ and therefore with $\beta_{q}(\bar{\lambda}_{q})\in \partialN E_{t}$ as claimed in {\small\bf D1-1-(a)}. We prove now the continuity of $\bar{\lambda}_{q}$ at $\bar{q}=\alpha_{p}(\mu_{p})$ ($\bar{q}=\alpha_{p}(\mu_{p})$ from now on).

First, from the proof of {\bf\small D1-1} one knows that $p$ is a smooth point of the boundary of a past cone $J^{-}(\gamma(\tau))$ entirely included in $E_{t}$. Moreover $\xi$ points strictly outwards to $J^{-}(\gamma(\tau))$ at $p$. It follows that for any sequence $q_j \rightarrow \bar{q}$ we have
$\liminf \{\bar{\lambda}_{q_j}\} \geq \bar{\lambda}_{\bar{q}}$. Indeed, let
$\tilde{\lambda}_{q_{j}}$ be the first $\lambda>0$ the orbit reaches the smooth boundary of $J^{-}(\gamma(\tau))$ near $p$.
Since  $\tilde{\lambda}_{q_{j}}\rightarrow -\mu_{p}$ and  $\bar{\lambda}_{q_{j}}\geq \tilde{\lambda}_{q_{j}}$ the claim follows.
We need to prove therefore that $\limsup \{\bar{\lambda}_{q_j} \} \leq \bar{\lambda}_{\bar{q}}$. Suppose instead that there is a sequence $q_j \rightarrow \bar{q}$ such that $\limsup
\{ \bar{\lambda}_{q_j} \} >\bar{\lambda}_{\bar{q}}+b$, for some $b>0$ and $b<a$. Since $E_{t}$
is closed (this is proved easily), it follows  that the piece of orbit $\{\beta_{\bar{q}}(\lambda),
0\leq \lambda\leq \bar{\lambda}_{\bar{q}}+b\}$ lies inside $E_{t}$. We
claim that, as a consequence, there are points $\beta_{\bar{q}}(\lambda)$, for
$\lambda>\bar{\lambda}_{\bar{q}}$ but arbitrarily close to
$\bar{\lambda}_{\bar{q}}$ lying in the interior of $E_{t}$. If not,
we would have that for $\lambda>\bar{\lambda}_{\bar{q}}$, the points
$\beta_{\bar{q}}(\lambda)$ must lie in $\partialN
E_{t}$. But by {\small\bf D1-1}, if a point in an orbit belongs to
$\partialN E_{t}$, then the points (in the orbit) near it and
in the direction opposite to $\xi$ are interior points to
$E_{t}$, which is a contradiction to the fact
$\beta_{\bar{q}} (\bar{\lambda}_{\bar{q}}) \in \partialN E_{t}$
Thus, the orbit $\beta_{\bar{q}}(\lambda)$ satisfies the following
properties:
\begin{enumerate}
\item $\beta_{\bar{q}}(\bar{\lambda}_{\bar{q}})\in \partialN
  E_{t}$,
\item $\beta_{\bar{q}}(\lambda)\in \Int(E_{t})$, for $0\leq
  \lambda<\bar{\lambda}_{\bar{q}}$,
\item there are points $\beta_{\bar{q}}(\lambda)\in \Int(E_{t})$,
  for $\lambda>\bar{\lambda}_{\bar{q}}$ but arbitrarily close to it,
\end{enumerate}

\n Les us show that these three facts together contradict {\small\bf D1-1}. We work now with the notation
$\alpha_{p}(\mu)=\beta_{\bar{q}}(\bar{\lambda}_{\bar{q}}+\mu)$ instead of the notation $\beta_{\bar{q}}(\lambda)$. Let $\mu_{1}>0$ 
be such that $\alpha_{p}(\mu_{1})$ belongs to the interior $E_{t}$. Let now
$\gamma(s)$, $s\geq 0$ be a future directed time-like geodesic, starting at $p$. Consider the orbits $\alpha_{\gamma(s)}(\mu)$, with $\alpha_{
\gamma(s)}(0)=\gamma(s)$ and $s>0$, but close to it. We
observe that $\gamma(s)\notin E_{t}$ (otherwise $p\in \Int(E_{t})$) and if $s>0$ is small enough then
$\alpha_{\gamma(s)}(\mu_{1})$ belongs to the interior of $E_{t}$.          
Thus we have  

\ben
\alpha_{\gamma(s)}(\mu_{1})\in \Int(E_{t}),\ \alpha_{\gamma(s)}(0)\in (\M^{+}\setminus E_{t}). 
\een

\n Since $\mu_1 >0 $ is as small as desired this immediately contradicts {\small\bf D1-1} and {\small\bf D1-2-(a)} is proved. 

Thus the map $q\rightarrow \beta_{q}(\bar{\lambda}_{q})$ (making $V_{p}$ smaller if necessary) is from $V_{p}$ into $\partialN E_{t}$. We have then $\lambda_{q}=\bar{\lambda}_{q}$ for $\lambda_{q}$ as defined in Definition \ref{D1}. Now, the argument that showed the continuity of $\bar{\lambda}_{q}$ at $q=\bar{q}$ shows the continuity of $\lambda_{q}$ at any point $q\neq \bar{q}$, namely {\small\bf D1-2-(b)}, and also {\small\bf D1-2-(c)}.\ep

\vspace{0.6cm}
\n {\it Proof of Proposition \ref{P3}}: The fact that the infinite Killing development is a smooth manifold
is seen as follows. The transition functions from $\K(V_{p})$ into
$\Int(E_{t})$ (on their domains of identification) according to identification {\small\bf D2-1}
 are trivially diffeomorphisms because they are given by 
\ben 
(q,\lambda)\rightarrow \beta_{q}(\lambda), 
\een
for $q\in V_{p}$ and $0<\lambda<\lambda_{q}$. Consider now the
transitions functions from $\K(V_{p_{1}})$ into
$\K(V_{p_{2}})$ (on their domains of identification) according to
the identifications {\small\bf D2-2}. We show that they are also
diffeomorphisms. First we show that the transition functions are one
to one and then we show that they are locally differentiable. Suppose
that $x_{1}=(q_{1},\lambda_{1})$ and $x'_{1}=(q'_{1},\lambda'_{1})$ in
$\K(V_{p_{1}})$ are identified to $x_{2}=(q_{2},\lambda_{2})$ in
$\K(V_{p_{2}})$ via {\small\bf D2-2}. Then, because
$\beta_{q_{1}}(\lambda_{q_{1}})$ and 
$\beta_{q'_{1}}(\lambda_{q'_{1}})$ must both be equal to
$\beta_{q_{2}}(\lambda_{q_{2}})$ it follows $q_{1}=q'_{1}$ and
$\lambda_{q_{1}}=\lambda_{q'_{1}}$. On the other hand 
\begin{gather*}
\lambda_{q_1} - \lambda_{1} = \lambda_{q_2} - \lambda_2,\text{ and } \lambda_{q'_1} - \lambda'_{1} =\lambda_{q_{2}} -\lambda_2
\end{gather*}

\n Thus $\lambda_{1}=\lambda'_{1}$ and therefore $x_{1}=x'_{1}$. This shows that
the transitions functions from $\K(V_{p_{1}})$ into
$\K(V_{p_{2}})$ (on their domains of identifications) are one to
one. We show now that they are locally differentiable. Suppose that
$x_{1}=(q_{1},\lambda_{1})\in \K(V_{p_{1}})$ and
$x_{2}=(q_{2},\lambda_{2})\in \K(V_{p_{2}})$, are identified
according to {\small\bf D2-2}. Then, we have
$\beta_{q_{1}}(\lambda_{q_{1}})=\beta_{q_{2}}(\lambda_{q_{2}})$. Let
$V$ be a smooth three-manifold (without boundary)
transversal to $\xi$ everywhere
satisfying $V \subset \K(V_{p_1}) \cap \Int(E_{t})$,
$V \subset \K(V_{p_2}) \cap \Int(E_{t})$
(both intersections under the natural identification {\small \bf D2-1}) 
and such that the orbit $\beta_{q_1}(\lambda)$ intersects $V$. Let
\ben
\hat{\varphi}_{1}: B_{1}\subset V_{p_{1}}\rightarrow
\K(V_{p_{1}}), 
\een
be the embedding satisfying $\hat{\varphi}_1(B_1) = V$. It is clear
that $B_1$ is an open neighbourhood of $V_{p_1}$ around $q_1$. As
a simple example $V$ could be chosen as the image of the graph 
\begin{eqnarray*}
\hat{\varphi}_{1}: B_{1}\subset V_{p_{1}} & \rightarrow & 
\K(V_{p_{1}}), 
 \\ 
\bar{q}_1 & \rightarrow &
\hat{\varphi}_{1}(\bar{q}_{1}) =(\bar{q}_{1},\lambda_{q_{1}}-\epsilon) 
\end{eqnarray*}

\n where $B_{1}$ is a sufficiently small open neighborhood of $V_{p_{1}}$
around $q_{1}$ and $\epsilon>0$ is a sufficiently small fixed number,
both chosen in such a way that
$V \equiv \hat{\varphi}_1 (B_1) \subset \K(V_{p_{2}})\cap \Int(E_{t})$

Since
$V \subset \K(V_{p_{2}})$, there exists a neighbourhood $B_2$
of $q_2$ in $V_{p_2}$ and an embedding $\hat{\varphi}_2 : B_2 
\rightarrow \K(V_{p_2})$ such that $\hat{\varphi}_2 (B_2) = V 
\subset \K(V_{p_2})$. Restricting $\hat{\varphi}_{1},
\hat{\varphi}_{2}$ to their images, we have two diffeomorphisms
\begin{gather*}
\varphi_{1}:B_{1}\rightarrow V,\ \ \varphi_{2}:B_{2}\rightarrow V.
\end{gather*}   
Consider now two open sets $\tilde{B}_1$ and $\tilde{B}_2$ defined
by
\n 
\begin{align*}
\tilde{B}_{1}=\{&(\bar{q}_{1},\lambda),\bar{q}_{1}\in B_{1},\\ 
& \lambda\in (\lambda(\hat{\varphi}_{1}(\bar{q}_{1}))-\lambda(\hat{\varphi}_{1}(q_{1}))+\lambda_{1}-\delta,
\lambda(\hat{\varphi}_{1}(\bar{q}_{1}))-\lambda(\hat{\varphi}_{1}(q_{1}))+\lambda_{1}+\delta)\},\\
\tilde{B}_{2}=\{&(\bar{q}_{2},\lambda),\bar{q}_{2}\in B_{2}, \\
& \lambda\in (\lambda(\hat{\varphi}_{2}(\bar{q}_{2}))-\lambda(\hat{\varphi}_{2}(q_{2}))+\lambda_{2}-\delta, \lambda(\hat{\varphi}_{2}(\bar{q}_{2}))-\lambda(\hat{\varphi}_{2}(q_{2}))+\lambda_{2}+\delta)\}
\end{align*}
\n where $\delta>0$ is chosen sufficiently small so that $\tilde{B}_1 \subset \K(V_{p_1})$
and $\tilde{B}_2 \subset \K(V_{p_2})$. The map
\ben
\phi:\tilde{B}_{1}\rightarrow \tilde{B}_{2}, 
\een
\n defined by $\phi(\bar{q}_{1},\lambda)=(\varphi_{2}^{-1}\circ
\varphi_{1}(\bar{q}_{1}),\lambda-\lambda(\hat{\varphi}_{1}(\bar{q}_{1}))+
\lambda(\hat{\varphi}_2(\varphi^{-1}_{2}(\varphi_{1}(\bar{q}_{1})))))$
is the transition function according to {\small\bf D2-2}
 restricted to
$\tilde{B}_{1}$ and is a smooth diffeomorphism onto its image
$\tilde{B}_{2}$.

The Hausdorff property of the abstract Killing development is seen as
follows. If $x_{1}=(q_{1},\lambda_{1})\in \K(V_{p_{1}})$ and
$x_{2}\in \Int(E_{t})$ are different points then either
$\lambda_{1}\geq \lambda_{q_{1}}$ or not, but if not then
$\beta_{q_{1}}(\lambda_{1})\neq x_{2}$. In either case it is
straightforward to find separating neighborhoods. Now, if
$x_{1}=(q_{1},\lambda_{1})\in \K(V_{p_{1}})$ and
$x_{2}=(q_{2},\lambda_{2})\in \K(V_{p_{2}})$ and different points
then either $\beta_{q_{1}}(\lambda_{q_{1}})\neq
\beta_{q_{2}}(\lambda_{q_{2}})$ or not, and if not then
$\lambda_{1}\neq \lambda_{2}$. Also in any of these possibilities it is straightforward to
find separating neighborhoods.

To see that the abstract Killing development is second countable use
that $\partialN E_{t}$ is a Liptschitz three-manifold, pick a
dense and countable set of points $\{p_{i}\}$ in $\partialN
E_{t}$ and over each point find a $V_{p_i}$ and construct
$\K(V_{p_{i}})$. Finally define the countable open subsets of
$\K(V_{p_{i}})$, $U_{ijklm}=B_{ij}\times (k/l-1/m,k/l+1/m)$ ($k,l$
naturals and $k/l>1/m$) where $B_{ij}$ is a countable basis of open
sets of $V_{p_{i}}$. The sets $\{U_{ijklm}\}$ together with a countable
basis of open sets of $\Int(E_{t})$ gives a countable basis for the
abstract Killing development.   \ep

\begin{figure}[h]\centering
\psfrag{TSd}{{\tiny $\partial \Sigma_d$}}
\psfrag{TSd2}{{\tiny $\partial \Sigma_{2d}$}}
\psfrag{TS2d}{{\tiny $\partial \Sigma_{d/2}$}}
\psfrag{TSigma}{{\tiny $\partial \Sigma$}}
\psfrag{Sr}{{\tiny $S_r$}}
\psfrag{Sr2}{{\tiny $S_{r+1}$}}
\psfrag{Sigmad}{{\tiny $\Sigma$}}
\psfrag{K}{{\tiny Killing vector}}
\psfrag{L}{{\tiny Light cones}}
\psfrag{E0}{{\tiny $\Ec$}}
\psfrag{Ei}{{\tiny $E_t$}}
\psfrag{D1}{{\tiny $\D_1$}}
\psfrag{D2}{{\tiny $\D_2$}}
\psfrag{D3}{{\tiny $\D_3$}}
\psfrag{U1}{{\tiny $\D_{-1}$}}
\psfrag{U2}{{\tiny $\D_{-2}$}}
\psfrag{U3}{{\tiny $\D_{-3}$}}
\psfrag{Lp2}{{\tiny $L_{2}$}}
\psfrag{B2}{{\tiny $B^2_{t}$}}
\psfrag{B3}{{\tiny $B^3_{t}$}}
\psfrag{Lm2}{{\tiny $L_{-2}$}}
\psfrag{gam}{{\tiny $\gamma(\tau)$}}
\psfrag{Cauchy}{{\tiny $\underline{\Sigma}_{\tilde{\tau}}$}}
\psfrag{Xi}{{\tiny $\xi$}}
\includegraphics[width=14cm,height=6cm]{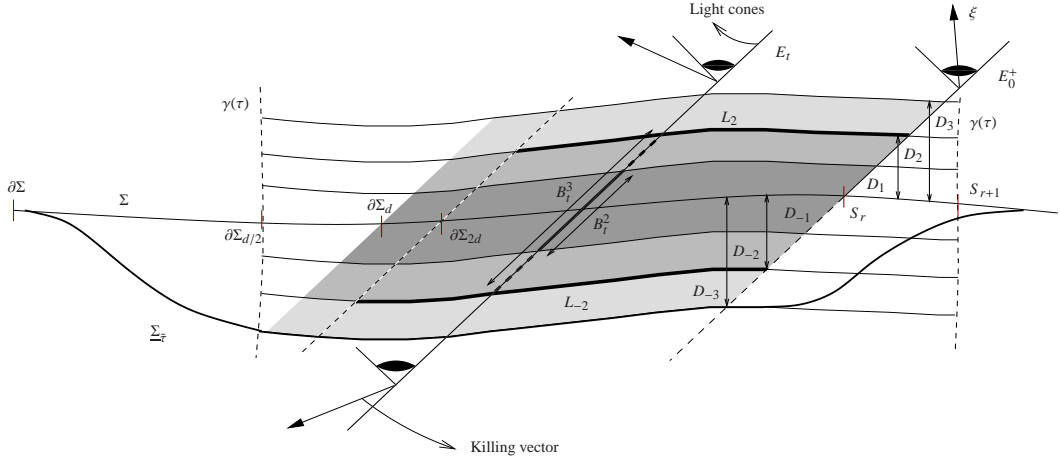}
\caption{Schematic figure illustrating the definitions of {\it slabs}
$\D_{-3}, \D_{-2}, \D_{-1}, \D_{1}, \D_{2}, \D_{3}$, layers
$L_{-2}, L_{2}$ and bands $B^{2}_t, B^{3}_t$. The initial Cauchy surface $\Sigma$,
the modified Cauchy surface $\underline{\Sigma}_{\tilde{\tau}}$
and the sets $\Ec$ and $E_{t}$ are also shown. The set $\D$ used in the text
is the union of all regions in grey in the figure.}
\label{Figure1}
\end{figure}

\vspace{0.6cm}
Before going into the proof of Proposition \ref{P4} we need to introduce some sets and their terminology. They are in fact simple regions of $\M$ although their precise definitions are somehow lengthy. The relevant sets to be used in the proof of the Proposition \ref{P4} are: the slabs $\D_{i},i=-3,\ldots,3$, the layers $L_{i},i=1,2$ and the bands $B^{i}_{t},i=2,3$. The graphic representation of the sets can be seen in Figure \ref{Figure1}. 

\vs
Given $p\in \IntSigma$ we consider the space-time (inextensible) timelike geodesic
$\gamma_{p}(\tau)(\subset \M)$ starting perpendicularly
to $p$ into the future and parametrized by proper time $\tau$.
Now, given $\Omega$ a compact region in $\IntSigma$, we define the tubular neighborhoods 
\ben
U(\Omega;\bar{\tau}_{1},\bar{\tau}_{2}) \defi\{\gamma_{p}(\tau),\ p\in
  \Omega,\bar{\tau}_{1}\leq \tau\leq \bar{\tau}_{2}\}
\een
We define also 
\begin{gather*}
\bar{\partial} U(\Omega;\bar{\tau}_{1},\bar{\tau}_{2})\defi\{\gamma_{p}(\tau),\ p\in
\Omega,\tau= \bar{\tau}_{2}\},\\
\underline{\partial} U(\Omega;\bar{\tau}_{1},\bar{\tau}_{2})
\defi \{\gamma_{p}(\tau),\ p\in \Omega,\tau=\bar{\tau}_{1}\}
\end{gather*}
For $\bar{\tau}$ small enough these are compact sets inside $\M$. We fix now $\tilde{\tau}>0$ such that
\begin{enumerate}
\item[{\small\bf P1.}] For every $p\in \Sigma_{d/2}\cap \Sigma^{I}(r+1)$, the geodesic $\gamma_{p}(\tau)$ is defined at least on the interval $[-3\tilde{\tau},3\tilde{\tau}]$. Moreover the map from $(\Sigma_{d/2}\cap \Sigma^{I}(r+1))\times [-3\tilde{\tau},3\tilde{\tau}]$ into $\M$, given by $(p,\tau)\rightarrow \gamma_{p}(\tau)$, is a diffeomorphism into the image. 

\item[{\small\bf P2.}] There is a Cauchy surface 
$\underline{\Sigma}_{\tilde{\tau}}$ for $\M$
containing the boundary $\underline{\partial} 
U(\Sigma_{d/2}\cap \Sigma^{E}(r+1);-3\tilde{\tau},3\tilde{\tau})$ and coinciding with $\Sigma$ outside a compact set in $\IntSigma$.

\item[{\small\bf P3.}]  The sets $U(\partial \Sigma_{d/2};-3\tilde{\tau},3\tilde{\tau})$ and $U(S_{r+1};-3\tilde{\tau},3\tilde{\tau})$ do not intersect the set
\ben
\bigg(D^{+}(\Sigma_{d})\cup J^{-}(\Sigma_{d})\bigg) \setminus \Ec
\een

\end{enumerate}
Then, given such $\tilde{\tau}$, we define the {\it slabs}
\begin{align*}
\D_{i} &\defi \bigg( D^{+}(\Sigma_{d})\cap U(\Sigma_{d/2}\cap\Sigma^{I}(r+1);0,i\tilde{\tau})\bigg)\setminus \Ec,\ i=1,2,3,\\ 
\D_{i} & \defi \bigg(J^{-}(\Sigma_{d})\cap U(\Sigma_{d/2}\cap \Sigma^{I}(r+1);i\tilde{\tau},0)\bigg)\setminus J^{-} (\Ec),
\ i=-1,-2,-3,\\
\D & \defi \D_{-3}\cup \D_{3}.
\end{align*}
Define the {\it layers} $L_{2}$ and $L_{-2}$ as
\begin{gather*}
L_{2}= (D^{+}(\Sigma_{2d})\cap \bar{\partial} U(\Sigma_{d/2}\cap\Sigma^{I}(r+1);0,2\tilde{\tau})\bigg)\setminus \Ec,\\
L_{-2}=\bigg(J^{-}(\Sigma_{2d})\cap \underline{\partial} U(\Sigma_{d/2}\cap \Sigma^{I}(r+1);-2\tilde{\tau},0)\bigg)\setminus J^{-} (\Ec)
\end{gather*}
Finally, suppose that $\W_t\Subset \M^{+}$ and let $E_{t}=J^{-}(\W_t)$. Then, define the bands $B^{i}_{t}$, $i=2,3$ and their upper and lower boundaries $\bar{\partial} B^{i}_{t}$, $\underline{\partial} B^{i}_{t}$
\begin{align*}
B^{i}_{t} & =(\partial ^{N} E_{t})\cap U(\Sigma_{d/2}\cap \Sigma^{I}(r+1);-i\tilde{\tau},i\tilde{\tau}),\\ 
\bar{\partial} B^{i}_{t} & =B^{i}_{t}\cap \bar{\partial} U(\Sigma_{d/2}\cap \Sigma^{I}(r+1),-i\tilde{\tau},i\tilde{\tau}),\\
\underline{\partial} B^{i}_{t} & =B^{i}_{t}\cap \underline{\partial} U(\Sigma_{d/2}\cap \Sigma^{I}(r+1);-i\tilde{\tau},i\tilde{\tau}) \\
B^{i \, \circ}_{t} & = B^{i}_{t} \setminus ( \bar{\partial} B^{i}_{t} \cup
\underline{\partial} B^{i}_{t} )
\end{align*}

\vs
\n {\it Proof of Proposition \ref{P4}}: We define $t_{*}$ as the supremum of the times $\bar{t}>0$ such that every Killing orbit $\beta_{q}(\lambda)$, $0 \leq \lambda< 2\bar{t}$, where $q\in L_{-2}\cup L_{2}$, and $\beta_{q}(0)=q$ lies inside
$\mbox{Int}(\D_{3}\setminus \D_{1})  \cup 
\mbox{Int}(\D_{-3}\setminus \D_{-1})$. Note that we are taking the range of $\lambda$ between $0$ and $2\bar{t}$ and not between $0$ and $\bar{t}$. 

We proceed now with the proof. Assume then {\small\bf H1} and {\small\bf H2}. We prove first {\small\bf C1}. We note two important observations concerning Killing orbits starting at $B^{2}_{t}$ that will be relevant for the discussion that follows. 
\begin{enumerate}
\item[{\small\bf O-1.}]\label{Obs-1} For any $p\in B^2_{t}$ the Killing orbit $\beta_{p}(\lambda)$, $\lambda>0$ remains inside $\Int(\D\setminus E_{t})$ until a first $\lambda$ when it reaches $\partial^{T} (D\setminus E_{t})\setminus B^{3}_{t}$. (The orbit cannot touch $B^{3}_{t}$ because $\xi$ points strictly inwards to $D\setminus E_{t}$ at $B^{3}_{t}$).
\item[{\small\bf O-2.}]\label{Obs-2} Because of {\small \bf O-1},
every pair of orbits $\beta_{p_{1}}(\lambda),\ \lambda\in (0,\lambda_{1})$ and $\beta_{p_{2}}(\lambda),\ \lambda\in (0,\lambda_{2})$ lying in $\Int(\D\setminus E_{t})$, with $p_{1},p_{2}\in B^{2}_{t}$  but different, do not intersect.
\end{enumerate} 
We prove now that $O_{[0,2t_{*}]} (B^{2}_{t}) \subset (D\setminus \Int(E_{t}))$. Assume that such is not the case and let $\bar{t}_{m}$ be
the minimum of the times $\bar{t}$, with $0<\bar{t}< t_{*}$ and such that
\ben
O_{[0,2 \bar{t}]}(B^{2}_{t}) \subset \M\text { and } O_{[2 \bar{t}]}(B^{2}_{t}) \cap \bigg(\M \setminus (D\setminus \Int(E_{t}))\bigg) \neq \emptyset
\een
Let $\bar{p}\in B^{2}_{t}$ be such that 
\ben
p=O_{[2\bar{t}_{m}]}(\bar{p})\in \bigg(\partial^{T} (D\setminus \Int(E_{t})) \setminus B^{3}_{t}\bigg) 
\een
where we are assuming that $p$ is not in $B^{3}_{t}$ because of {\small\bf O-1}. Let $\gamma(\tau)$, $\tau\in [0,1]$, be a past directed null geodesic inside $B^{2}_{t}$ starting at $\bar{q}\in \bar{\partial} B^{2}_{t}$ and ending at $\bar{p}$. Then $O_{[2\bar{t}_{m}]}(\gamma(\tau))$ is a past directed null geodesic starting at $q=O_{[2\bar{t}_{m}]}(\bar{q})$ and ending at $p$. But by definition of $\bar{t}_{m}$, it is $\bar{t}_{m}<t_{*}$ and therefore it must be $q\in \Int(\D_{-3}\cup \D_{3})$. Because of {\small\bf O-1} the geodesic $O_{[2\bar{t}_{m}]}(\gamma(\tau))$ cannot intersect $B^{3}_{t}$. Therefore by the definition of $D$ it must be $p\in \underline{\Sigma}_{\tilde{\tau}}$. That this is an impossibility is seen as follows. First note that
\ben
\partial^{T} O_{[0,2\bar{t}_{m}]}(B^{2}_{t})=B^{2 \,
\circ}_{t} \cup O_{[2\bar{t}_{m}]}(B^{2 \, \circ}_{t})\cup O_{[0,2 \bar{t}_{m}]} (\underline{\partial} B^{2}_{t}) \cup O_{[0,2\bar{t}_{m}]} (\bar{\partial} B^{2}_{t})
\een
and that because of {\small\bf O-2} the union on the right hand side is disjoint. Second we claim that inextensible past directed time-like geodesics $\Gamma(\tau)$, $\tau\geq 0$, starting at the point $p$ (found before) at $\tau=0$ must remain inside $\Int (O_{[0,2 \bar{t}_m]}(B^{2}_{t}))$ (for $\tau>0$) until a first $\tau=\bar{\tau}$ when it reaches
\be\label{SD}
B^{2 \, \circ}_{t} \cup O_{[0,2\bar{t}_{m}]} (\underline{\partial} B^{2}_{t}) \cup O_{[0,2\bar{t}_{m}]} (\bar{\partial} B^{2}_{t})
\ee
Indeed if instead there is such a $\Gamma(\tau)$ and $\bar{\tau}>0$ with $\Gamma(\bar{\tau})\in O_{[2\bar{t}_{m}]}(B^{2 \, \circ}_{t}))$ then $O_{[-2t_{m}]}(\Gamma(\tau))$, with $\tau$ near $\bar{\tau}$, would be a past directed time-like geodesic inside $E_{t}$ and crossing $B^{2 \, \circ}_{t}$ at $O_{[-2\bar{t}_{m}]}(\Gamma(\bar{\tau}))$ which is not possible as $B^{2}_{t}\subset \partialN E_{t}=\partial ^{N} J^{-}(\W_t)$. Thus any past directed time-like geodesic $\Gamma$ starting at $p$ would eventually touch (\ref{SD}). But because the set (\ref{SD}) lies to the future of $\underline{\Sigma}_{\bar{\tau}}$ and $p\in \underline{\Sigma}_{\bar{\tau}}$ we obtain an impossibility. We have thus proved that $O_{[0,2t_{*}]} (B^{2}_{t}) \subset (D\setminus \Int(E_{t}))$ as we wanted. 

Because of this and because of {\small\bf O-2} we claim that we can construct a natural differentiable map from $O_{(0,2t_{*})}(B^{2 \, \circ}_{i}))$
into $K(E_{t},2t_{*})^{\circ}$, which is actually an isometry. In other words we claim that we have naturally $O_{(0,2t_{*})}(B^{2 \, \circ}_{t}))\Subset K(E_{t},2t_{*})^{\circ}$. Roughly speaking the isometry can be explained in the following terms: We can think of $B^{2}_{t}$ both 
as a set in $\M$ or as a set in $K(E_{t},2t_{*})$ then the map identifies Killing orbits in $\M$ starting at points in $B^{2 \, \circ}_{t}$, as a set in $\M$, with Killing orbits in $K(E_{t},2t_{*})$ starting in $B^{2 \, \circ}_{t}$, but now as a set inside $K(E_{t},2t_{*})$. In precise terms, the map is defined as follows. Let $o$ be a point in $ O_{(0,2t_{*})}(B^{2 \, \circ}_{t})$. We will define the map in a neighborhood of it. We can write $o=O_{[\bar{t}]}(p)$ with $0<\bar{t}<2t_{*}$, and with $p\in B^{2 \, \circ}_{t}$. Both, $\bar{t}$ and $p$, are unique because of {\small\bf O-2}. Let $\mu_{p}<0$, $q=\alpha_{p}(\mu_{p})$, $V_{p}\subset E_{t}$ and $\K(V_{p})$ be as in Definition \ref{D1}. Then for every point $o'$ in a neighborhood of $o$ there are $q(o')\in V_{p}$ and $\lambda(o')$ (the correspondences $o'\rightarrow q(o')$ and $o'\rightarrow \lambda(o')$ being smooth) such that $o'=\beta_{q(o')}(\lambda_{o'})$. The map $o'\rightarrow (q(o'),\lambda (q(o'))\in \K(V_{p})\subset K(E_{t},2t_{*})$ is the desired map. Following the identifications in Definition \ref{D1} (which define $K(E_{t},2t_{*})$), it is simple to see that the map we defined is indeed independent of the choice of $V_{p}$. 

With this identification in mind we consider now the set 
\ben 
\Omega_{t+2t_{*}}\defi(O_{[0,2t_{*})}(B^{2}_{t}) \cap \Sigma)\cup \overline{\Omega_{t}}, 
\een
as a set inside $K(E_{t},2 t_{*})^{\circ}$, where $\Omega_{t}= \Int(E_{t})\cap \Sigma$. We claim that $\Omega_{t+2t_{*}}$ is a 
Cauchy surface of the subset $F$ of $K(E_{t},2t_{*})^{\circ}$
\ben
F=F_{1}\cup F_{2}\cup F_{3},
\een
where
\begin{align*}
F_{1}&=E^{+}_{t},\\
F_{2}&=O_{[0,2t_{*})}((\partial^{N}E^{+}_{t})\setminus B^{2}_{t}),\\
F_{3}&=O_{[0,2t_{*})}(B^{2}_{t})\cap \M^{+}
\end{align*}
and where to define $F_{2}$ and $F_{3}$ as subsets of $K(E_{t},2t_{*})^{\circ}$ we are using the identification constructed before. To see the claim note first that $\Omega_{t}$, which is a subset of $\Omega_{t+2t_{*}}$ is a Cauchy surface for $E^{+}_{t}$.  Then by noting that every inextensible past directed causal curve in $E_{t}$ starting at a point in $E_{t}^{\circ}$ cannot reach $\partial^{N}E_{t}$, conclude that every inextensible past directed causal curve in $F$ starting at a point in $F_{2}\cup F_{3}$ must either first reach $\partial^{N} E^{+}_{t}$ or eventually reach $\Omega_{t+2t_{*}}\setminus \Omega_{t}$. The claim follows.

Now as  $\Omega_{t+2t_{*}}\Subset \IntSigma$
we have $F\Subset \M^{+}$. We obtain therefore $K(E^{+}_{t},t_{*}) \subset (F\cup O_{[0,2t_{*}]}(B^{2}_{t}))$,
but $F\Subset \M$ and $O_{[0,2t_{*}]}(B^{2}_{t})\Subset \M$, thus $K(E^{+}_{t},t_{*})\Subset \M$, which proves {\small\bf C1}.

\vs
We now show {\small\bf C2}. Suppose {\small\bf C2} is false. Since by {\small\bf H2}\  $T(\partial \Sigma,2d) \cap E_{t} = \emptyset$, there exists $0<\bar{t}<t_{*}$ such that $J^{-} ( K(E_{t},\bar{t}))$ intersects $T(\partial\Sigma,2d)$ and no smaller $0<\bar{t}<t_{*}$ has this property. Let $p \in \partial T(\partial\Sigma,2d)\cap J^{-} ( K(E_{t},\bar{t}))$ and let $\gamma_{p}(\tau)$ be the future directed null geodesic on $\partialN J^{-} ( K(E_{t},\bar{t}))$ starting at $p$. Consider $\gamma_{p}(\tau'_j)$ where
$\tau'_j\rightarrow \infty$
is a divergent sequence. Then we know
\begin{enumerate}
\item $(J^{-}(\gamma_{p}(\tau'_j))\cap \Sigma) \subset
(J^{-} ( K(E_{t},\bar{t}))\cap \Sigma)\subset \Sigma_{2d}$,
\item $p$ is a smooth point of $\partialN J^{-}(\gamma_{p}(\tau'_j))$,
\item $\partialN J^{-}(\gamma_{p}(\tau'_j))\cap \Sigma$ is tangent 
to $\partial \Sigma_{2d}$ at $p$.
\end{enumerate} 

\n Thus a standard comparison of mean curvatures implies that the 
expansion $\bar{\theta}^{+}(p)$ of

\n $\partialN J^{-}(\{\gamma_{p}(\tau'_j)\})\cap \Sigma$ at $p$ is less
or equal than the expansion $\theta^{+}(p)$ of $\partial \Sigma_{2d}$
at $p$, hence negative. By the Rauchadury equation the foliation of null geodesics of
$\partialN J^{-}(\{\gamma_{p}(\tau'_j)\})$ must develop a focussing point
along $\gamma_p(\tau)$ in a parametric affine parameter less than a fixed
value depending on $\theta^{+}(p)$. This contradicts the fact that $\tau'_j\rightarrow
\infty$ and that $\gamma_{p}(\tau)$ has no focal points between
$\tau=0$ and $\tau=\tau'_j$.   

\vs
We show now {\small\bf C3}. We want to prove
\be\label{CP2}
J^{-}(K(E^{+}_{t},t_{*}))\cap \M^{+}=J^{-}(\W_{t+t_{*}})\cap \M^{+}
\ee
The inclusion of the right hand side into the left hand side follows directly because $\W_{t+t_{*}}\Subset K(E_{t}^{+},t_{*})$. We prove now the inclusion of the left hand side into the right hand side. Let $p\in J^{-}(K(E^{+}_{t},t_{*}))\cap \M^{+}$. Then there is $q\in K(E^{+}_{t},t_{*})$ and a future causal curve $\gamma_{1}$ joining $p$ to $q$. If $q\in E^{+}_{t}$ then we are done as $E^{+}_{t}=J^{-}(\W_{t}) \cap \M^{+}$ and therefore there is a future causal curve $\gamma_{2}$ joining $q$ to a point in $\W_{t}\subset \W_{t+t_{*}}$. Thus $\gamma_{2}\circ \gamma_{1}$ (the concatenation of $\gamma_{1}$ and $\gamma_{2}$) is a future causal curve joining $p$ to a point in $\W_{t+ t_{\star}}$. Hence $p$ belongs to 
the right hand side of (\ref{CP2}). If instead $q\in K(E^{+}_{t},t_{*})\setminus E^{+}_{t}$ then $q=O_{[\bar{t}]}(q')$ where $0<\bar{t}\leq t_{*}$ and $ q'\in \partialN E^{+}_{t}$. Then there is a null geodesic $\gamma_{2}$ inside $\partialN E^{+}_{t}$ starting at $q'$ and eventually becoming a null geodesic of $\partialN \W_{t}$. Therefore $O_{[\bar{t}]}(\gamma_{2})$ is a future null geodesic starting at $q$ and eventually becoming a null geodesic of $\partialN \W_{t+t_{*}}$. Therefore the curve $O_{[\bar{t}]}(\gamma_{2})\circ \gamma_{1}$ is a future causal curve joining $p$ to a point in $\W_{t+t_{\star}}$. \ep

\vs
 \n {\it Proof of Theorem \ref{ThmExterior}:} Suppose that $\Sigma^{T}\setminus \Omega_{\infty}\neq \emptyset$ (recall we are using $\Omega_{\infty}=\Sigma\cap E^{+}_{\infty}=\Sigma\cap (\cup E^{+}_{i}$)). Let $p\in \partial^{T}(\Sigma^{T}\setminus \Omega_{\infty})$ (where the topological boundary of $\Sigma^{T}\setminus \Omega_{\infty}$ is taken as it were a set in $\Sigma^{T}$). Then $\xi(p)$ is time-like and future directed. Let $\{q_{j}\}\subset \Sigma^{T}\cap \Omega_{\infty}$ be a sequence approaching $p$, namely $q_{j}\rightarrow p$. Then, there is $\infty>\lambda_{0}>0$ such that the piece of orbit $\beta_{q_{j}}(\lambda),\ \lambda\in (0,\lambda_{0}]$ lies in  $\M^{+}$ for all $j$. Note that for every $j$ there is $i(j)$ such that $q_{j}\in \Omega_{i(j)}$ (and therefore that $q_{j}\in E^{+}_{i}$) for all $i\geq i(j)$. We claim that the piece of orbit above also lies in $E^+_{\infty}$. To see that observe that to leave $E^{+}_{\infty}$  it must first leave $E^{+}_{i}$ for all $i\geq i(j)$. That means that for every $(i,j)$, $i\geq i(j)$ there are $\lambda_{j_{i}}<\lambda_{j_{i+1}}<\lambda_{0}$ such that $\beta_{q_{j}}(\lambda),\ \lambda\in (\lambda_{j_{i}},\lambda_{j_{i+1}})$ lies inside $E^{+}_{i+1}\setminus E^{+}_{i}$ and that $\beta_{q_{j}}(\lambda_{j_{i}})\in \partial^{N} E^{+}_{i}$ and $\beta_{q_{j}}(\lambda_{j_{i+1}})\in \partial^{N}
E^+_{i+1}$. Then, because $E^{+}_{i+1}=J^{-}(K(E^{+}_{i},t_{*}))\cap \M^{+}$, the piece of orbit $\beta_{q_{j}}(\lambda),\ \lambda\in (\lambda_{j_{i}},\lambda_{j_{i+1}})$ must lie inside $O_{[0,t_{*}]}(\partial^{N} E^{+}_{i})\cap \M^{+}\subset E^{+}_{i+1}$. Therefore it must be $\lambda_{j_{i+1}}-\lambda_{j_{i}}\geq t_{*}$. It follows from here that, given $j$, then $\lambda_{j_{i}}\rightarrow \infty$ as $i\rightarrow \infty$. Thus it must be $\lambda_{0}=\infty$ which is a contradiction, and the orbit $\beta_{q_{j}} (\lambda), \lambda \in (0,\lambda_0]$ lies in $E^+_{\infty}$.
Finally we observe that because $\beta_{q_{j}}(\lambda)$, $\lambda\in [0,\lambda_{0}]$ is a time-like curve starting at $q_{j}$ and ending at a point in $E^{+}_{\infty}$, and therefore ending in one of the $E^{+}_{i}$'s, then if $q_{j}$ is sufficiently close to $p$ the point $p$ will lie in the  interior of $\Omega_{\infty}$ which is against the hypothesis.\ep

\section{Static Killing initial data}
\label{Static}

\subsection{Background and definitions}

We start with the notion of {\it static Killing initial data}.

\begin{Definition}
\label{statickid}
A static Killing initial data (static KID) set $\Da$ is 
a KID satisfying the staticity equations
\begin{align}
& N d \Y + 2 \Y \wedge Z = 0,  \label{static1} \\
&  \Y \wedge d \Y = 0, \label{static2}
\end{align}
where $Z \defi d N + K (Y, \cdot \, )$.
\end{Definition}

In a static KID, consider the open
set $\SigmaY \defi  \Sigma^{\circ}
\setminus \{ \Y = 0 \}$. By the Fr\"obenius theorem, the distribution
${\Y}^{\bot}$ is integrable. More precisely,
each point $p \in \Sigma^Y$ is contained in a unique, maximal,
arc-connected, injectively immersed $(m-1)$-dimensional, orientable
submanifold $\L_{\alpha}$ orthogonal to $\Y$.
The collection of $\{ \L_{\alpha} \}$ is a foliation of $\Sigma^Y$.
The staticity equation (\ref{static1}) and (\ref{kid1}) imply
\begin{eqnarray}
\lambda d \Y + \Y \wedge d \lambda = 0.
\label{integ}
\end{eqnarray}
As a consequence
of this equation, if $\lambda =0$ (resp. $\lambda>0$, $\lambda <0$)
at any point $p \in \L_{\alpha}$ then $\lambda =0$ (resp.
$\lambda>0$, $\lambda <0$) everywhere on $\L_{\alpha}$. To see this,
consider any path $\gamma(s)$ contained in $\L_{\alpha}$. Contracting
(\ref{integ}) with $Y$ and $\dot{\gamma}$ we obtain the ODE
\begin{equation*}
\frac{d\lambda (s)}{ds} = Q(s) \lambda(s),
\end{equation*}
where $Q(s)$ is smooth
and $\lambda(s) \defi  \lambda (\gamma(s))$. The claim follows.

As discussed in the Introduction, the aim of this part of the work
is to show that Killing prehorizons of the exterior region are necessarily embedded.
Killing prehorizons are immersed null hypersurfaces where the Killing vector
is null and tangent (hence also normal). Thus, their intersection with $\Sigma$
must correspond to those leaves ${\cal L}_{\alpha}$
where $\lambda$ vanishes identically. Since we are interested only
on horizons of the exterior region or, more precisely, on
horizons than can be reached from the exterior, timelike region, we
adopt the following definition.

\begin{Definition}
\label{horizon}
A horizon $\H_{\alpha}$
is  a leaf of the foliation $\{\L_{\alpha}\}$ of $\SigmaY$
which intersects the topological boundary $\tbd \SigmaT$.
\end{Definition}

Any two points $p_1, p_2$ on  a fixed leaf $\L_{\alpha}$ admit
transverse sections (i.e. smooth connected curves that are transverse to all
the leaves they intersect) $\Gamma_1$ and $\Gamma_2$ to the foliation 
$\{{\cal L}_{\alpha} \}$
and a smooth diffeomorphism $\phi: \Gamma_1 \rightarrow \Gamma_2$ such that
for any leaf $\L_{\beta} \in \{{\cal L}_{\alpha} \}$ one has
$\phi (\L_{\beta} \cap \Gamma_1 ) = \L_{\beta} \cap \Gamma_2$
(this property is the so-called {\it transverse
uniformity of foliations}, see e.g. Theorem 3, p. 49 in \cite{Neto}).
As a consequence, any horizon $\H_{\alpha}$ is fully contained
in $\tbd \SigmaT$.

 Since $\lambda$ vanishes on a horizon, 
$d \lambda$ is necessarily a normal one-form
to $\H_{\alpha}$. Consequently
there exists a scalar function $\kappa_{\alpha}$ on $\H_{\alpha}$,
called the {\it surface gravity},
satisfying $d \lambda = 2 \kappa_{\alpha} \Y$ on ${\cal H}_{\alpha}$.
It is 
also convenient to introduce a scalar on 
$\Sigma$ defined as 
\begin{equation}
\Ione \defi \frac{1}{4} |d \Y|_g^2 - 2 |Z|_g^2. \label{Iuno}
\end{equation}
An alternative expression for $\Ione$ on the set $\{ N \neq 0 \}$
(in particular on $\{ \lambda > 0\}$) is obtained from 
the staticity equation (\ref{static1}), which gives
$d \Y = - \frac{2}{N} \left ( \Y \wedge Z \right )$. Squaring this
and inserting
into the definition of $\Ione$ yields 
\begin{align}
\Ione = \frac{2}{N^2} ( - \lambda |Z|_g^2 - \langle \Y,Z \rangle_{g}^2 )
\quad \quad \mbox{on} \quad \{ N \neq 0 \}. 
\label{expIone} 
\end{align}

Passing to the Killing development (which exists in a neighbourhood
of any point $p \in \H_{\alpha}$ since $N |_p \neq 0$), it follows
from standard properties of Killing horizons that $\kappa_{\alpha}$
is constant
on each horizon $\H_{\alpha}$ (see e.g.
\cite{Wald} p. 334 for a derivation in four dimensions which is, in fact,
valid in any dimension). 

The following lemma relates $\Ione$ to the surface
gravity on horizons.
\begin{Proposition}
\label{surface}
$\Ione = - 2 \kappa_{\alpha}^2$ on $\H_{\alpha}$. 
\end{Proposition}

\n {\it Proof}: From (\ref{expIone}), it suffices to show that 
$\kappa_{\alpha} = \frac{2 \langle \Y, Z\rangle _g}{N}$.
Contracting (\ref{kid1})
with $Y$ gives $K(Y,Y) = - \frac{1}{2N} Y ( |Y|_g^2 )$ which inserted in $Z$
(see Definition \ref{statickid}) gives, on $\H_{\alpha}$,
\begin{align}
\frac{2 \langle \Y, Z \rangle_g }{N}  =
  \frac{1}{N^2} Y \left ( N^2  - |Y|_g^2 
\right ) = 
\frac{1}{N^2} Y (\lambda )  = 2 \kappa_{\alpha}.
\end{align}
\ep 
\vs 

Horizons with non-zero surface gravity have properties qualitatively
different to horizons with vanishing surface gravity. The following definition
is standard.
\begin{Definition}
\label{dege}
A horizon $\H_{\alpha}$ is degenerate if $\kappa_{\alpha}=0$
and non-degenerate if $\kappa_{\alpha} \neq 0$.
\end{Definition}

Points where the Killing vector vanishes correspond, at the initial data
level, to points $p \in \Sigma$ satisfying $N |_p = Y |_p = 0$.
Such points are called {\it fixed points}. 
The following lemma is well-known in static four-dimensional
spacetimes. At the initial
data level, it has been proved in four dimensions in \cite{AlbertoTesis}.
We include a proof for $m-$dimensional static KIDs in Appendix A. 
\begin{lemma}
\label{fixed}
$\Ione < 0$ on any fixed point $p \in \tbd \SigmaT$.
\end{lemma}

\vs 
In this part of the paper we intend to work directly at the initial data level. This has the advantage that
no assumption on well posedness of the matter model needs to be made. Nevertheless, we still
require the null energy condition to hold. The following definition translates the
standard spacetime definition into the initial data setting.
\begin{Definition}
\label{null}
A Killing initial data set $\Da$ satisfies the null energy
condition if and only if
\begin{eqnarray*} 
\Tau (w,w )  - 2 J (w) |w|_g + \rho |w|_g^2  \geq 0 
\end{eqnarray*}
for any vector $w \in T_p \Sigma$ and $p \in \Sigma$.
\end{Definition}

\subsection{The statements of the main results: Theorem \ref{Thm2} and Corollary \ref{Cor_2} }
\label{theresults2}
Our main result in this second part of the paper is the following.

\begin{Theorem}
\label{Thm2}
Let $\Da$ be an asymptotically flat static Killing initial data set satisfying the null energy condition.
Suppose that $\partial \Sigma$ (if non-empty)  does not intersect $\Sigma^T$. Then, each degenerate horizon is 
an embedded manifold and compact.
\end{Theorem}

An immediate Corollary of Theorems \ref{ThmExterior} and \ref{Thm2} is
\begin{Corollary}
\label{Cor_2}
Let $\Da$ be an asymptotically flat static Killing initial data set  with well
posed
matter model satisfying the null energy condition. Suppose that
$\partial \Sigma$ (if non-empty) is future outer trapped. Then each degenerate horizon 
is an embedded manifold and compact.
\end{Corollary}

\begin{Remark} It may be possible to prove, directly from the techniques that we developed here, a version of Theorem \ref{Thm2} also for stationary data and not just static. We will not enter into such problem here however.
\end{Remark}

\subsection{Volume monotonicity along ``optic" congruences of geodesics}

In this section, we will assume that the datum $\Da$ is static (Definition \ref{statickid}). 

\vs 
The Killing development of a static KID is static in the sense
that the Killing vector $\xi$ is hypersurface orthogonal (see
Lemma 3 in \cite{CarrascoMars2008}). Static spacetimes 
necessarily satisfy $\Einbf (\xi,X)=0$, where $X$ is
any vector  field orthogonal
to $\xi$. In terms of the quantities $(\rho,J,\Tau)$ defined
by (\ref{Projection}), this implies
\begin{eqnarray}
\Tau ( Y, \cdot \,)   = N J + \frac{J (Y )}{N}  \Y - \rho \Y, 
\quad \quad \mbox{on } \{ N \neq 0 \} \subset \Sigma. \label{tauY}
\end{eqnarray}
In addition to $g$, $\SigmaT$ can be endowed with two further metrics:
the so-called {\it quotient metric} 
\begin{equation}
h \defi
g + \frac{1}{\lambda} \Y \otimes \Y
\label{defih}
\end{equation}
and the  {\it optic metric} 
\begin{equation}
\label{defihbar}
\hbar  \defi \frac{1}{\lambda} h.
\end{equation}
Consider the spacetime $(\SigmaT \times \mathbb{R} ,\gS)$ 
with metric
\begin{equation}
\gS = - V^2 dt^{\prime \, 2} + h \label{3+1}
\end{equation}
where $ V \defi  + \sqrt{\lambda}$. Equation (\ref{integ}) implies that $\lambda^{-1} \Y$ is closed
on open sets where $\lambda$ does not vanish, in particular on 
$\SigmaT$. Consequently there exists, locally, a function
$\zeta$ such that $\Y = - \lambda d\zeta$.
The coordinate transformation $t = t' - \zeta$
brings the metric $\gD$ (see (\ref{development})) 
into $\gS$. This shows that 
the spacetimes $(\SigmaT \times \mathbb{R},\gD)$ and
$(\SigmaT \times \mathbb{R},\gS)$
are locally isometric. They are also globally isometric if 
$\lambda^{-1} \Y$ is exact on $\SigmaT$.

Since the data on $\{t'=0 \}$ in the metric (\ref{3+1}) is a totally geodesic
static KID, it satisfies the constraint equations (\ref{C1})-(\ref{C2})
and the KID equations (\ref{kid1})-(\ref{kid2}) with the substitutions
$g \rightarrow h$, $N
\rightarrow V$, $Y \rightarrow 0$ and $K \rightarrow 0$. With the 
definitions $\rhoh \defi  V^{-2} \Einbf(\xi,\xi)$ and
$\Tauh (v,w) \defi  \Einbf (v,w)$, with $v,w$ tangent to $\{t'=0\}$, these
equations read
\begin{align}
& \Hess_h  V = V
\left ( \Rich - \Tauh + \frac{1}{m-1}
\left ( \tr_{h} \Tauh - \rhoh \right ) h \right ), \label{Rich} \\
& \Deltah V = V \left ( \frac{\tr_h \Tauh}{m-1} + \frac{m-2}{m-1} \rhoh
\right ), \label{laplacian}
\end{align}
where $\Deltah$ is the Laplacian of $h$ and
$\Rich$ its Ricci tensor. 
Using $\xi = N n + Y$ and (\ref{Projection}), it is straightforward to
relate $\rhoh$, $\Tauh$ to the datum $\Da$, as follows
\begin{eqnarray}
\rhoh = \rho - \frac{1}{N} J ( Y), \quad \quad
\Tauh = \Tau  + \frac{1}{\lambda} \left (
\frac{J(Y)}{N} - \rho \right ) \Y \otimes \Y. \label{matter}
\end{eqnarray}

The following Proposition characterizes the null energy condition of the initial
data set in terms of the geometry associated to $h$. 
\begin{Proposition}
\label{energy}
$\Da$ satisfies the null energy condition if and only if
\begin{eqnarray*}
\Tauh \left ( \wh, \wh \right ) + \rhoh  |\wh |^2_h \geq 0
\end{eqnarray*}
for any vector $\wh \in T_p \SigmaT$ and $\forall p \in \SigmaT$.
\end{Proposition}
{\it Proof}: This Proposition can be proved easily by passing to the
Killing development of $\SigmaT$. For a direct proof 
on the initial data set, consider any vector vector $\wh$ and
define  $w \defi  \wh + \frac{A}{V} Y$, where $A \defi  
|\wh|_{h} + \frac{\langle Y,\wh \rangle_g}{V}$. The $g$-norm of $w$ is calculated to
be $|w|^2_g = \frac{N^2 A^2}{\lambda}$. 
A straightforward computation which uses (\ref{matter})
and (\ref{tauY}) gives
\begin{eqnarray*}
\Tau (w,w )  - 2 J (w) |w|_{g} + \rho |w|_g^2 = \Tauh ( \wh,\wh) + \rhoh |\wh|^2_{h}. 
\end{eqnarray*}
Since transformation $\wh \rightarrow w$ 
is invertible (with inverse $\wh = w -\frac{|w|_g}{N} Y$), the Proposition follows from
Definition \ref{null}. \ep

Expression (\ref{Rich}) determines the Ricci tensor of $h$ in terms
of $V$ and its derivatives. A similar expression can be obtained 
for the Ricci tensor of $\hbar$, denoted by $\Richbar$. We write $\nablah$ for the
covariant derivative of $h$ and $\nablahbar$ for the covariant derivative
of $\hbar$.
\begin{Proposition}
The Ricci tensor of $\hbar$ takes the following form
\begin{eqnarray}
\Richbar = (m-1) \frac{1}{V} \Hess_h V
- (m-1)  \frac{|\nablah V|^2_{h}}{V^2} h  +   \rhoh  h +
\Tauh. \label{Rhbar}
\end{eqnarray}
\end{Proposition}
{\it Proof}: The general expression for the change of Ricci tensor under
a conformal rescaling $\hbar = e^{2f} h$ is
\begin{eqnarray*}
\Richbar = \Rich + (2- m) \left ( \Hess_h f -
df \otimes df \right ) - \left ( \Deltah f + (m -2)
|\nablah f|^2_h \right ) h.
\end{eqnarray*}
Putting $f = - \mbox{ln} (V)$ and inserting (\ref{Rich}) and (\ref{laplacian}), the 
Proposition follows.\ep

The following proposition is well-known \cite{Weyl} 
and explains the reason
of calling $\hbar$ the {\it optic} metric.
\begin{Proposition}
\mbox{}
\begin{enumerate}
\item 
Let $\gamma(t)$, $t \in [t_0,t_1]$ be a geodesic segment
in $(\SigmaT,\hbar)$ parametrized by $\hbar$-arc-length. Select
$c \neq 0$, define
\begin{eqnarray*}
\tau(t) = \tau_0 + \int_{t_0}^{t} c^{-1} V^2(\gamma(t)) dt
\end{eqnarray*}
and denote by $t(\tau)$ its
inverse (which obviously exists). Then
the curve $(\gamma(t(\tau)), t(\tau))$, $\tau \in [\tau_0,\tau(t_1)]$
is an affinely
parametrized null geodesic segment
in $(\SigmaT \times \mathbb{R} , \g_S)$ and its
tangent vector $v$ satisfies
$\g_S (v, \xi) = - c$.
\item Conversely, let $(\gamma(\tau), t(\tau))$, $\tau \in [\tau_0,
\tau_1]$ be an
affinely parametrized null geodesic segment in $(\Sigma \times \mathbb{R},\gS)$
 with tangent vector
$v$. Define $c = - \g_S (v,\xi)$ (which 
is obviously constant along the geodesic) and define $\tau(t)$
as the inverse of 
\begin{eqnarray*}
t (\tau) \defi t_0 + \int_{\tau_0}^{\tau} \frac{c}{V^2 (\gamma(\tau))} d\tau.
\end{eqnarray*}
Then, the curve $\gamma(\tau(t))$, $t \in [t_0, t(\tau_1)]$
is a geodesic segment in $(\SigmaT,\hbar)$ parametrized by $\hbar$-arc-length.
\end{enumerate}
\end{Proposition}
Consider a smooth, oriented hypersurface $S$ embedded
in $\SigmaT$ and let $\m$ and $\overline{\m}$ be positively
oriented normal vectors, of unit length respectively in $h$ and $\hbar$ (they are
obviously related by $\overline{\m}= V \m$). Let $\gam$ (resp. $\gambar$)
denote the induced metric on $S$ inherited from $h$ (resp. $\hbar$).
The following fact is well-known
and straightforward.
\begin{Proposition}
With the notation before, the second fundamental form $\chi$
of $S$ with respect to
 $\m$ in the metric $h$ and the second fundamental
form $\overline{\chi}$ of $S$ with respect to $\overline{\m}$
in the metric $\hbar$ are related by
\begin{eqnarray*}
\overline{\chi} =  \frac{\chi}{V} 
- \frac{\m(V)}{V^2} \gam.
\end{eqnarray*}
\end{Proposition}
Squaring in their respective metrics and taking traces
the following expressions follow,
\begin{eqnarray}
& & \overline{\theta} = V \theta - (m-1) \m(V), \label{thetas}\\
& & |\overline{\chi}|^2_{\gambar} = V^2 | \Pi |^2_{\gam} +
\frac{V^2}{m-1} \left (
\theta - \frac{(m-1)}{V} \m(V) \right )^2, \nonumber
\end{eqnarray}
where $\theta \defi  \tr_{\gam} \chi$, 
$\overline{\theta} \defi  \tr_{\gambar} \, \overline{\chi}$ and
$\Pi$ is the trace-free part of $\chi$ (in the metric $\gam$).
The expression above
for $\overline{\theta}$ and the
Ricci tensor of $\hbar$ give rise to the following monotonicity formula.
\begin{Proposition}
\label{monotonicity}
Let ${\cal F}$ be a congruence of geodesics in 
$(\SigmaT,\hbar)$ parametrized by arc-length. Assume that the
tangent vector $\overline{\m}$ to this congruence
is orthogonal to
a collection of smooth hypersurfaces 
$\{\SS_t \}$. Then, the trace $\theta$ of the second fundamental
form $\chi$ of $\SS_t$ with respect to $\m$ satisfies
\begin{eqnarray}
\overline{\m} \left (\frac{\theta}{V} \right )
       + |\Pi |^2_{\gam}+ \frac{1}{m-1} \theta^2 + \rhoh
+ \Tauh \left ( \m,\m\right ) = 0. \label{mhtthh}
\end{eqnarray}
In particular, if the null energy condition is satisfied in $(\SigmaT,h)$ then 
\begin{eqnarray*}
\overline{\m} \left ( \frac{\theta}{V} \right ) \leq - \frac{1}{m-1} \theta^2 \leq 0.
\end{eqnarray*}
\end{Proposition}
{\it Proof}: The focusing equation for geodesics (see e.g. \cite{Chavel})
is
\begin{eqnarray}
\overline{\m} \left ( \overline{\theta} \right )
 + |\overline{\chi}|^2_{\gambar}
+ \Richbar (\overline{\m},\overline{\m} ) = 0.
\label{focussing}
\end{eqnarray}
The term $\Richbar (\overline{\m},\overline{\m})$ can be directly
evaluated from (\ref{Rhbar}):
\begin{equation}
\Richbar(\overline{\m},\overline{\m})  = (m-1) V \, \Hess_h V \left (\m,\m \right )
- (m-1) |\nablah V|^2_{h} + V^2 \left ( \rhoh + \Tauh (\m,\m ) \right ). \label{Rhbarmtmt}
\end{equation}
In order to evaluate the term $\overline{\m} (\overline{\theta})$
in (\ref{focussing}), the $h$-acceleration 
$\nablah_{\m} \m $ is needed. 
Since $\overline{\m}$ is geodesic and affinely parametrized
we have $\nablahbar_{\overline{\m}}
\overline{\m}=0$, which becomes, after applying
the transformation law for metric connections under conformal rescalings,
\begin{equation}
\nablah_{\m} \m = \frac{1}{V} \m(V)
\m - \frac{1}{V} \nablah (V). \label{nablahmh}
\end{equation}
We then have, from (\ref{thetas}),
\begin{align}
\overline{\m} (\overline{\theta}) & =
\overline{\m} (V) \theta + V \, \overline{\m} ( \theta)
- (m-1) V \left [
\left \langle \nablah_{\m} \m, \nablah V \right \rangle_h
+(\Hess_h V) \left ( \m,\m \right ) \right ]  \nonumber \\
& = \overline{\m} (V) \theta + V \, \overline{\m} ( \theta)
+ (m-1) \left ( |\nablah V|^2_h - \m(V)^2 -
V (\Hess_h V) \left ( \m,\m \right ) \right ). \label{mhtthht}
\end{align}
Inserting (\ref{Rhbarmtmt}) and (\ref{mhtthht}) into 
(\ref{focussing}), the terms in the Hessian of $V$ cancel out. A
simple rearrangement gives (\ref{mhtthh}). 
The last claim follows from Proposition \ref{energy}. \ep



\subsection{On the volume of horizons of asymptotically flat static KIDs}

Recall that $\Sigma^{\infty}$ is the AF end of $\Da$.
The decay (\ref{decay}) implies that $(\Sigma^{\infty},h)$ is also 
asymptotically flat. Let $S_r$ and $\Sigma^I(r)$ be defined as in Section
\ref{background} and define $\SigmaT_I \defi \Sigma^I(r) \cap \Sigma^T$. 
We start by showing that $\SigmaT_I$ is complete in the metric 
$\hbar$.
\begin{lemma}
\label{completeness}
Assume that $\SigmaT$ does not intersect $\partial \Sigma$. Then,
the Riemannian manifold $(\SigmaT_I,\hbar)$ is complete and has $S_r$
as its only boundary.
\end{lemma}
\n {\it Proof}: First we make a couple of comments on the structure of the metric $\bar{h}$ around (I) a point on a horizon and (II) a fixed point.

(I) Consider a point $p$ lying on a 
horizon $\H_{\alpha}$ and choose a foliated chart $(V_p, \{ x^A,z\})$ near
$p$ adapted to the foliation $\{\cal L_{\alpha} \}$.
This means
that, in these coordinates, $V_p = \Omega \times (-\delta,\delta) $,
where $\Omega$ is a domain
on $\mathbb{R}^{m-1}$ and $\delta >0$. The coordinate $z$
takes values in $(-\delta,\delta)$  and $\{x^A\}$ ($A,B = 1, \cdots, m-1$)
takes values in $\Omega$.
The intersection of any leaf 
$\L_{\alpha}$ with $V_p$ is a collection (possibly empty)
of sets of the form $\Omega \times \{ z_i \}$ (called plaques)
where $\{ z_i \}$ is a countable set. Since $Y$ is
$g$-orthogonal to the plaques, we can choose, without
loss of generality, the coordinate chart so that
the metric $g$ takes the form
$g = F^2(z,x^C) dz^2 + \hat{g}_{AB} (z,x^C) dx^A dx^B$, 
where $F>0$ and $\hat{g}_{AB}$ is positive definite. Furthermore, we
can assume that $p = \{ 0 \}$. Let $P$ be the 
smooth positive function 
on $V_p$ such that
$\Y = N P \nu$ where $\nu$ is $g$-unit and orthogonal to the plaques. This
implies $\lambda = N^2 ( 1- P^2)$.  Inserting all this into (\ref{integ})
yields, after a straightforward calculation,
\begin{equation*}
\left ( 1 - P^2 \right )
 \left ( - \frac{\partial_A F}{F} + \frac{\partial_A N}{N} 
\right ) - \frac{1 + P^2}{P} \partial_A P  = 0, 
\quad \mbox{for all \hspace{2mm}} z \in (-\delta,\delta)
\end{equation*}
which, upon integration, implies the existence of a smooth function $U$
on $V_p$, constant on every plaque (i.e. $U(z)$), such that 
\begin{equation*}
\frac{N}{F} \frac{1 - P^2}{P} = U.
\end{equation*}
Using this into the definition of $\hbar$ (\ref{defihbar}) gives
\begin{equation}\label{OMH}
\hbar = \frac{dz^2}{P^2 U(z)^2}  + \frac{1}{N \, F \, P \, U(z)}
\hat{g}_{AB} dx^A dx^B.
\end{equation}

Note that from $\lambda |_p =0$ we have $U(z=0)=0$. Also, as $U$ is
differentiable, we have (on $\Sigma^{T}$) $0<U\leq c|z|$ (where $c>0$ is a constant) near $p$.

(II) Let now $p$ be a fixed point. Then we know from (\ref{Iuno}) and Lemma
\ref{fixed} that $Z|_p \neq 0$. This in turn implies $dN |_p \neq 0$
(see Definition \ref{statickid}). Thus, there exists a neighbourhood
$V_p$ of $p$ where $N$ can be taken as a coordinate. Without loss
of generality, we can choose a coordinate system in $V_p$ so that
$g = \tilde{F}^2(N,x^C) dN^2 + \tilde{g}_{AB} (N,x^C) dx^A dx^B$
where $\tilde{F}>0$ and $\tilde{g}_{AB}$ is positive definite.
By the definition of $\hbar$, we have
\begin{eqnarray}\label{OMFP}
\hbar \geq \frac{1}{\lambda} g \geq \frac{1}{N^2} g.
\end{eqnarray}
Note that $N|_{p}=0$ and that, once more, $N$ is a coordinate in a differentiable coordinate system. 

We are ready to prove completeness of $(\Sigma^{T}_I,\bar{h})$. Assume by contradiction that $\bar{h}$ is not complete. Let $\gamma$ be an incomplete $\bar{h}$-geodesic not ending at $S_{r}$. Then $\gamma$, as a curve over $\Sigma$ accumulates (although not necessarily converging to) a point $p$ on a horizon or a fixed point $p$. From the structure of the metric $\bar{h}$ found around such points in (\ref{OMH}) and (\ref{OMFP}), respectively, one readily deduces that the $\bar{h}$-length of $\gamma$ must be infinite which is a contradiction. \ep

\vs 

On $\SigmaT_{I}$ define $\B$, $t > 0$, as  
the $\hbar$-ball of center $S_r$ and radius $t$, 
\ben
B(t,S_{r})=\{ p \in \Sigma^T_I,\mbox{dist}_{\hbar} (p, S_r) < t\}. 
\een
The boundary component
\begin{equation*}
\bd \B \defi \tbd \B \setminus S_r
\end{equation*}
is  the set of points
lying at $\hbar$-distance $t$ to $S_r$. Outside the cut locus ${\cal C}$
this set of points is a smooth hypersurface. We want to consider
the (m-1)-Hausdorff measure of $\bd \B$ {\it in the metric $h$}, which 
we denote by $|\bd \B|_h$. 
The following lemma gives an upper bound for $|\bd \B|_h$.

\begin{lemma}
\label{upperbound}
Let $S_r$ be the coordinate sphere of radius $r$ in $\Sigma^{\infty}$
and assume that the $h$-mean curvature with respect to the ingoing unit vector  is negative everywhere. Assume that $\SigmaT$ does not intersect $\bd \Sigma$  and
let $|S_r|_h$ be the  $(m-1)$-volume of $S_r$ in the metric $h$.
Then, $|\bd \B|_h \leq |S_r|_h$ for all $t > 0$.
\end{lemma}

\n {\it Proof}: On $\SigmaT_I$ consider the  congruence
${\mathcal{F}}$ of geodesics minimizing the $\hbar$-distance to $S_r$. An immediate consequence of Lemma \ref{completeness}
is that each geodesic
in ${\mathcal F}$ has an end-point in $S_r$.
For any $p\in \Sigma^T_I$
outside the cut locus ${\mathcal{C}}$ of the distance function (which has zero measure 
\cite{MantegazzaMennucci}) the function
$t(p)=\mbox{dist}_{\hbar}(p,S_r)$ is smooth and in there the level sets of $t$
are smooth hypersurfaces. In other words, if $p \in \Sigma^T_I
\setminus \mathcal{C}$
then, around $p$, $\partial \Bp$ is a smooth hypersurface.
Let $p$ be such a point and let $\gamma_{p}(t)$ be the length
minimizing segment that starts at $S_r$ and ends at $p$. 
We define
the function $\theta$ on $\SigmaT \setminus {\mathcal{C}}$ as the
$h$-mean curvature of $\partial \Bp$
at $p$ in the direction of $\gamma'_{p}(t(p))$. Note  that
the mean curvature is with respect to $h$ and not $\hbar$,
but that the congruence ${\mathcal{F}}$ is with respect to $\hbar$ and not
$h$.

Now, from Proposition \ref{monotonicity} 
we have the monotonicity
\begin{equation*}
\overline{\m} \left ( \frac{\theta}{V} \right )
\leq -\frac{\theta^{2}}{m-1}\leq 0.
\end{equation*}
Since $\theta|_{S_r}<0$ we conclude that $\theta<0$ on
$\SigmaT \setminus {\mathcal{C}}$. Denoting by $\bm{\eta}_{h} (p)$
the volume-form of $\partial \Bp$ at $p \in \SigmaT \setminus {\cal C}$,
the first variation $(m-1)$-volume gives
\begin{equation*}
\overline{\m} (\bm{\eta}_h) = V \m (\bm{\eta}_h ) = V \theta <0.
\end{equation*}
This proves $|\bd \B|_h \leq |S_r|_h$.\ep


\vs
We analyze now the interplay between the $(m-1)$-volume
of horizons in the static KID and the $(m-1)$-volume of the $\hbar-$geodesic spheres 
$\bd \B$. 


Let $\H_{\alpha}$ be a horizon and let $\m$ be 
be one of the two possible normal vector fields to $\H_{\alpha}$. 
For every point $q\in \H_{\alpha}$ consider the $g$-geodesic $\gamma_{q}(s)$
starting at $q$ with velocity $\m(q)$ and parametrized with
arc-length. Let $\Omega \subset \H_{\alpha}$ be open and connected
with smooth and compact boundary in $\H_{\alpha}$.  
\begin{Definition}
Let $\H_{\alpha}$ be a horizon. We say that $\H_{\alpha}$ is 
isolated on $\Omega$ in
the direction of $\m$
if for some $\bar{s}$ small, the set (tubular
neighborhood of $\Omega$) 
\begin{equation*}
U_{\m}(\Omega,\bar{s})=\{\gamma_{q}(s),q\in \Omega, 0 < s < \bar{s}\},
\end{equation*} 
is contained in $\SigmaT$ and does not intersect any  horizon.
A horizon $\H_{\alpha}$ is isolated if there exists an
exhaustion $\{\Omega_i\}$ of $\H_{\alpha}$ such that
$\H_{\alpha}$ is  isolated on $\Omega_{i}$ in both normal
directions.
\end{Definition}
Since $d \lambda \neq 0$ everywhere on a 
non-degenerate horizon, it follows that non-degenerate horizons
are necessarily isolated.
\begin{Proposition} 
\label{upperboundareahorizons}
Let $\H_{\alpha}$ be an isolated horizon in the direction of
$\m$ over $\Omega$. Then 
\begin{eqnarray*}
\liminf_{t \rightarrow \infty}  | U_{\m}(\Omega,\bar{s}) \cap \bd \B |_h
\geq |\Omega|_g.
\end{eqnarray*}
where $|\Omega|_g$ is the $g$-(m-1)-volume of $\Omega$.
\end{Proposition}
{\it Proof}: We need several definitions first.
\begin{enumerate}
\item At every point $p\in U_{\m}(\Omega,\bar{s})$, let $\m$
be the tangent of the geodesic $\gamma_q(s)$
passing through $p$. Choose $(m-1)$ vector fields $\{e_1, \cdots
e_{m-1}\}$ on $U_{\m} (\Omega,\bar{s})$ such that
$\{e_{1},\cdots, e_{m-1},\m\}$ is an oriented $g$-orthonormal
basis. Let $\{ \omega^1, \cdots, \omega^m \}$
be the corresponding dual basis. Define then the $(m-1)$-form 
\begin{eqnarray*}
\omega= \omega^1 \wedge \cdots \wedge \omega^{m-1}.
\end{eqnarray*}
\item For every $0<\tilde{s}<\bar{s}$ define the surface 
\begin{eqnarray*}
S(\Omega,\tilde{s})=\{\gamma_{q}(\tilde{s}),q\in \Omega\},
\end{eqnarray*}
and its one-sided tubular neighbourhoods 
\begin{eqnarray*}
U^+ (\Omega,\tilde{s})= \{\gamma_{q}(\tilde{s}),q\in \Omega,
\tilde{s} \leq s < \bar{s}\}, 
\quad
U^- (\Omega,\tilde{s})= \{\gamma_{q}(\tilde{s}),q\in \Omega,
0 < s \leq \tilde{s}\}.
\end{eqnarray*}  
\end{enumerate}

\n Now, for every $\tilde{s}$ there is $t_{0}(\tilde{s})$ such that
if $t>t_{0}(\tilde{s})$ then $U^+ (\Omega,\tilde{s})
\subset \B$,
namely $U^+(\Omega,\tilde{s})$ lies in the interior of the 
$\hbar$-metric ball
$\B$.  For such $t$ define $B (t,\tilde{s})$
as the connected component of 
$U_{\m}(\Omega,\tilde{s})\cap \B$, containing 
$U^+(\Omega,\tilde{s})$. Then $\tbd
(B(t,\tilde{s}) \setminus U^+(\Omega,\tilde{s}))$ consists of: 
\begin{enumerate}
\item $S(\Omega,{\tilde{s}})$, 
\item an interior component that we will denote $\SS_{t}(\Omega)$ which
is in fact equal to a component of $U^-(\Omega,\tilde{s})\cap \partial B(t,\tilde{s})$, and,
\item a domain inside the $(m-1)$-surface $\{\gamma_{g}(s),q\in \partial
\Omega,0 < s < \tilde{s}\}$.
\end{enumerate}
Since the metric $h$ is related to $g$ by (\ref{defih}) their volume forms are related by
\begin{eqnarray*}
\bm{\eta}_h = ( 1 + \frac{|Y|_g^2}{\lambda} ) \bm{\eta}_g.
\end{eqnarray*}
Consequently
\begin{eqnarray*}
|\SS_{t}(\Omega)|_{h} \geq |\SS_{t}(\Omega)|_g.
\end{eqnarray*}

\n On the other hand we have
\begin{eqnarray*}
|\SS_{t}(\Omega)|_g \geq \int_{\SS_{t}(\Omega)} \omega = |S(\Omega,\tilde{s}) |_g
+\int_{(B(t,\tilde{s}) \setminus U^+(\Omega,\tilde{s}))} d \omega.
\end{eqnarray*}
\n Integration by parts is justified (for almost all $t$) because the distance function is Lipschitz and therefore of
bounded variation \cite{EvansGariepy} (indeed it is semiconcave and therefore a $H^{2,1}$ function
\cite{MantegazzaMennucci}. But now, as $\tilde{s}\rightarrow 0$ and $t>t_{0}(\tilde{s})
\rightarrow \infty$, the first term on the right hand side approaches
$|\Omega|_g$ and the second converges to zero. Since
obviously 
$| U_{\m}(\Omega,\bar{s}) \cap \bd \B |_h \geq |\SS_{t} (\Omega)|_{h}$ we conclude 
\begin{eqnarray*}
\liminf_{t \rightarrow \infty} | U_{\m}(\Omega,\bar{s}) \cap \bd \B |_h
\geq \liminf_{t \rightarrow \infty} |\SS_{t}(\Omega)|_{h}  \geq |\Omega|_g.
\end{eqnarray*}
\ep

\begin{Proposition}
\label{closure}
Assume that $\Sigma^{T} \cap \partial \Sigma = \emptyset$.
Let $\{ H_{\alpha} \}_{{\alpha} \in {\cal J}}$ be any collection of horizons
in an asymptotically flat static KID $\Da$. Let $ H =
\bigcup_{\alpha \in {\cal J}} \H_{\alpha}$ be its union. Then $\overline{H}
\setminus H$ is either empty or consists only of fixed points.
\end{Proposition}
{\it Proof}: The proof is by contradiction. We will assume that
there exists $p \in \overline{H} \setminus H$ which is not
a fixed point and we will show that $\lim_{t \rightarrow \infty} |\bd \B|_h = + \infty$, which contradicts the upper bound found in Lemma
\ref{upperbound}.

Let $p$ be such a point. Since $p$ is non-fixed
($Y|_p \neq 0$), there exists a unique leaf $\L_{\beta}$ containing 
$p$. Since
$H \subset \tbd \SigmaT$ (recall that a horizon is fully contained
in $\tbd \SigmaT$) and the latter is topologically closed, it follows
that $p \in \tbd \SigmaT$, so in fact $\L_{\beta}$ is a horizon $\H_{\beta}$.
By hypothesis, this horizon is not in the original collection 
$\{ H_{\alpha} \}_{\alpha \in {\cal J}}$. Consider a foliated chart 
$V_p$ of $p$ in $\SigmaY$ as in the proof of Lemma \ref{completeness}.
Without loss of generality, we can assume
that the foliated chart is centered at ${\cal H}_{\beta}$, i.e.
that the plaque $\Omega \times \{0\} \subset {\cal H}_{\beta}$. Also
without loss of of generality we assume that $\overline{\Omega}$  is compact
with smooth boundary. By definition of horizon, there
exists a sequence of points $p_i \rightarrow p$ with $p_i \in \SigmaT$
(in particular $\lambda (p_i)>0$). Moreover, since $p \in \overline{H} \setminus
H$, there must exists a sequence of plaques in $H$ converging to $\Omega \times \{ 0 \}$.
These two facts together imply the existence
of two sequences $a_i \rightarrow 0$, $b_i \rightarrow 0$, $-\delta < a_i < b_i
< \delta$ such that
\begin{enumerate}
\item $\Omega \times (a_i,b_i) \subset \SigmaT$,
\item $\Omega \times \{a_i\} \in \tbd \SigmaT$,
\item $\Omega \times \{b_i\} \in \tbd \SigmaT$.
\end{enumerate}

\vs
\begin{figure}[h]
\centering
\psfrag{z}{\tiny $z$}
\psfrag{z0}{\tiny $z=0$}
\psfrag{a1}{\tiny $a_1$}
\psfrag{a2}{\tiny $a_2$}
\psfrag{b1}{\tiny $b_1$}
\psfrag{b2}{\tiny $b_2$}
\psfrag{t1}{\tiny $t_1$}
\psfrag{t2}{\tiny $t_2$}
\includegraphics[width=12cm,height=7cm]{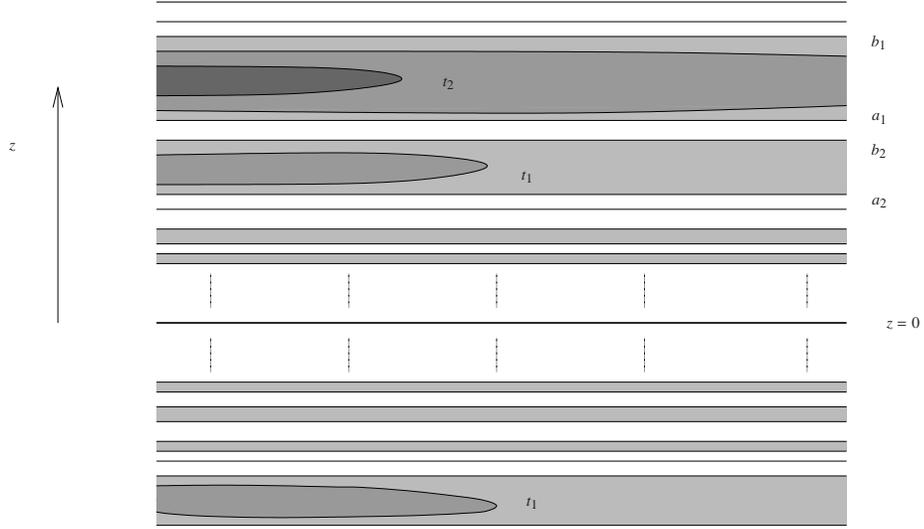}
\caption{Foliated neighbourhood of $\Omega$. The grey regions lie in 
$\Sigma^T$. Schematic plots of the balls $B(t,S_r)$ are shown for 
two values of $t$, namely $t_1$ and $t_2$ satisfying $t_1 < t_2$. As
$t$ increases the balls increase and approach the boundaries
of $\Sigma^T$. Two values of the sequences $\{a_i\}$ and $\{b_i\}$
defined in the main text are also shown. In the case of the figure $C(t_1) =0$
and $C(t_2)=1$. }
\label{Figure2}
\end{figure} 

Let $\H^+_{i}$ be the horizon containing the plaque
$\Omega \times \{b_i\}$, and
$\H^-_{i}$ be the horizon containing the plaque
$\Omega \times \{a_i\}$. Both horizons are isolated on $\Omega$ (the
first one in the direction of decreasing $z$ and the second
one towards increasing $z$).  For each $t$ sufficiently large, let
$C(t)$ be the maximum of $i \in \mathbb{N}$ such that,  
 $\forall j \leq i$, $\Omega \times (a_j,b_j) \cap \B$
is non-empty and has more than one connected component.
By Proposition \ref{upperboundareahorizons}, 
each one of the pieces
$\bd \B \cap (\Omega \times (a_i,b_i))$, $i \leq C(t)$
contributes to the total $(m-1)$-volume 
$|\bd \B|_h$ essentially with an amount of at least $2 |\Omega|_g$.
More precisely, for fixed $\epsilon > 0$,
there exists $t_0(\epsilon)$ such that for $t > t_0(\epsilon)$ 
\begin{eqnarray*}
|\bd \B|_h  \geq 2 C(t) \left ( |\Omega|_g - \epsilon \right )
\end{eqnarray*}
Since $C(t) \rightarrow \infty$ when $t \rightarrow \infty$, 
we obtain  $\lim_{t \rightarrow \infty} |\bd \B|_h = + \infty$ 
and hence a contradition to Lemma \ref{upperbound}.
\ep

We can now prove Theorem \ref{Thm2}.

\vs 
\n {\it Proof of Theorem \ref{Thm2}:} We first show that any degenerate horizon $\H_{\alpha}$ is compact.
$\Ione$ vanishes on $\H_{\alpha}$ and hence also on its closure.
Assume that there is a point $p \in \overline{ \H}_{\alpha} \setminus 
\H_{\alpha}$. $p$ must be a fixed point by Proposition \ref{closure}. However,
since $\H_{\alpha} \in \tbd \SigmaT$, it follows $p \in \tbd \Sigma^T$. Lemma \ref{fixed} gives
$\Ione < 0$, which gives a contradition. Hence $\H_{\alpha}$ is 
topologically closed. Closed leaves in foliations
are necessarily embedded (see e.g. Theorem 5, p. 51
in \cite{Neto}). Moreover, since $\H_{\alpha}$ is contained in the compact set
$ \Sigma^I(r)$, $\H_{\alpha}$ is also compact. \ep

The previous results prove not only that a degenerate
horizon cannot approach itself indefinitely, but also that two or more such
horizons cannot wrap on themselves indefinitely. More
precisely, we have the following Corollary.

\begin{corollary}
Assume that $\SigmaT \cap \bd \Sigma = \emptyset$.
Then all horizons are isolated. Moreover $|\bigcup_{\alpha} \H_{\alpha}|_g <
|S_r|_h$.
\end{corollary}
{\it Proof:} Assume that $\H_{\alpha}$ is a non-isolated horizon, hence
necessarily degenerate. 
Since $\H_{\alpha}$ is not isolated, 
there exists an open set $\Omega \subset \H_{\alpha}$
with compact, smooth boundary in $\H_{\alpha}$ such that for all
$\bar{s}>0$, the tubular neighbourhood
\begin{equation*}
U(\Omega,\bar{s})=\{\gamma_{q}(s),q\in \Omega, -\bar{s}<
s< \bar{s}\}, 
\end{equation*} 
intersects another horizon $\H_{\beta}$. Selecting a sequence
$\bar{s}_i \rightarrow 0$, we have a collection of horizons 
$\H_{\beta_i}$ which approach $\H_{\alpha}$. 
By Theorem  \ref{Thm2} $\H_{\alpha}$ is embedded.
It follows that at least one of the
${\cal H}_{\beta_i} \neq \H_{\alpha}$. Consider the collection ${\cal A}$
of all $\{ \H_{\beta_i} \}$ different from $\H_{\alpha}$. It follows
that the set $H \defi \cup_{\cal A} \H_{\beta_i} $ is not closed, 
as its closure contains $\Omega$. This contradicts Proposition
\ref{closure}. Thus $\H_{\alpha}$ is isolated.
The last statement is a direct 
consequence of Proposition \ref{upperboundareahorizons}
and Lemma \ref{upperbound}. \ep


%
%
%

\section{Uniqueness of static, vacuum, asymptotically flat initial
data sets with outer trapped boundary}
\label{uniq}

The results of the previous sections allow us to prove a uniqueness theorem
for asymptotically flat static KID with an outer trapped boundary. 

The most powerful method of proving uniqueness of static black holes
is the so-called {\it doubling method} of Bunting and Masood-ul-Alam
\cite{BuntingMasood}. The framework
where this method applies involves asymptotically flat KID such that
the exterior region $\SigmaT$ where the Killing vector is timelike 
has a topological boundary which is a compact, embedded $C^0$  manifold without boundary
(see \cite{Chrusciel1999} for details in the vacuum case).
More  generally, the method applies also to settings where the Cauchy boundary
of $\SigmaT$ is a compact, embedded $C^0$  manifold without boundary (the
Cauchy boundary is defined as the set of points lying in the Cauchy completion
of the set but not in the set itself). The need of using Cauchy completions comes
from the fact that a point $p \in \tbd \SigmaT$ may be accessible from both 
sides from within $\Sigma^T$. In this setting $\tbd \SigmaT$ may not be 
a $C^0$ manifold but the Cauchy boundary, denoted by $\partial^C \SigmaT$
may still be a topological manifold. We will see an example of this behavior 
later.

A possible strategy for proving a uniqueness theorem for static KID with an outer trapped boundary 
is to reduce the problem to a black hole uniqueness theorem. This suggests the following
definition (c.f. \cite{CarrascoMars2011, AlbertoTesis}): {\it
an  asymptotically flat KID $\Da$ (possibly with boundary) is a {\it black hole static initial data set} 
if the Cauchy  boundary $\partial^C \mbox{Int} (\Sigma^T)$
of $\SigmaT$ is a topological manifold without boundary and compact.}

In agreement with the discussion above, we
 will also say that a {\it matter model satisfies the 
static black hole uniqueness
theorem} if  there exists a class of asymptotically flat
static spacetimes $\{ (M_{\mathfrak{a}},\bm{g_{\mathfrak{a}}})\}$ 
depending an a finite
(and usually small) number of parameters determined from the asymptotic 
form of the metric and matter fields 
and such that any {\it black hole static Killing initial data set}
$\Da$ has the property that 
$(\Sigma^T,g,K)$ can be isometrically embedded in
some  $\{ (M_{\mathfrak{a}},g_{\mathfrak{a}})\}$ 
within this class (where isometrically embedded is in the sense
of spacetime initial data sets).

As example of matter models satisfying the static black hole theorem
we have vacuum or
electrovacuum in four spacetime dimensions 
(see
\cite{Chrusciel1999}, \cite{ChruscielTod} ,\cite{ChruscielGalloway} and references therein).

An interesting 
consequence of the results in the previous chapters is
that the static black hole uniqueness theorem can be extended to static,
asymptotically flat KID with outer trapped boundary.
More precisely
\begin{Theorem}
\label{uniqueness}
Let $\Da$ be a static, $m$-dimensional ($m \geq 3$) 
asymptotically
flat Killing initial data satisfying the following assumptions:
\begin{itemize} 
\item[{\small\bf A1}.] The matter model is well posed and satisfies the null
energy condition.
\item[{\small\bf A2.}] The matter model satisfies the 
static black hole uniqueness theorem.
\item[{\small\bf A3}.] $\Sigma$ has outer trapped boundary, i.e. $\partial \Sigma$ is
compact and $\theta^{+} (\partial \Sigma) < 0$.
\end{itemize}
\vs
Then $(\Sigma^T,g,K)$ can be isometrically embedded in
some $(\M_{\mathfrak{a}},\bm{g_{\mathfrak{a}}})$ within the black hole uniqueness
class.
\end{Theorem}

\begin{Remark}
In the particular case of vacuum, this result takes the form 
of Theorem \ref{uniqueness vacuum} given in the Introduction.
\end{Remark}

We first recall a well-known property 
of fixed points of Killing
vectors in spacetimes of arbitrary dimension.
\begin{lemma}
\label{KF}
Let $(\M,\g)$ be a spacetime with a Killing vector $\xi$. Let $p$
be a fixed point of $\xi$, i.e $\xi(p)=0$ and let $F \defi \frac{1}{2} 
d \bm{\xi} $ (where $\bm{\xi} = \g (\xi,\cdot \, )$) be
the so-called {\it Killing form}
of $\xi$. Define $W_p \defi \{ v \in T_p \M ;
F|_p (v, \cdot \,)=0\} $. Then $p$ lies in smooth, totally geodesic embedded
submanifold $S_p$ of dimension $k = \mbox{dim} (W_p)$ ($p$
is an isolated fixed point if $k=0$). Moreover, $S_p$
is spacelike, null or timelike depending on whether $W_p$ is spacelike,
null or timelike.
\end{lemma}
 
\n {\it Proof of Theorem \ref{uniqueness}}:  From Theorem  \ref{ThmExterior}
 it follows that $\Sigma^T$ does not intersect
$\partial \Sigma$. By Theorem \ref{Thm2} each degenerate
horizon of $\Sigma^T$ is compact and embedded. The same
is true for non-degenerate horizons. Consider any point
$p$ in a horizon. Then, near $p$ the
Cauchy boundary $\partial^C \Sigma^T$ either coincides
with $\tbd \Sigma^T$ (if $\Sigma^T$ only
lies to one side of $\tbd \Sigma^T$ at $p$) or with two
copies of $\tbd \Sigma$ (if $p$ can
be accessed from  both sides
within $\Sigma^T$). In either case, the Cauchy completion is a smooth
manifold near $p$. 

It only remains to analize the fixed points $p \in \partial
\Sigma^T$. We know from the proof of Lemma \ref{fixed}
in Appendix A that $dN |_p \neq 0$ and that
$d \Y |_p =
\frac{2b}{Q} ( dN|_p \wedge X )$ for some 
$b,Q >0$ and $X \in T^{\star}_p \Sigma$  which is both unit and orthogonal
to $dN|_p$. Moreover, $b^2 < Q^2$ from (\ref{IbQ}) in the proof of Lemma \ref{fixed}.
Let us now view $p$ as a point in $\M$. It is clear that $p$ is 
a fixed point for the Killing vector $\xi$.
The Killing form $F$ at $p$ is easily evaluated to be (c.f. Definition 3 
in \cite{CarrascoMars2008})
\begin{eqnarray*}
F|_p  =  (dN  \wedge \nn) |_p + \frac{b}{Q} ( dN|_p \wedge X ),
\end{eqnarray*}
where $\nn$ is the future directed unit normal one-form to $\Sigma$ in $\M$.
Since the one-form $(\nn|_p + b Q^{-1} X)$ is
timelike (from $Q^2 > b^2$), it follows
that $dN_p$ and $\nn|_p + b Q^{-1} X$ span a timelike two-plane. From the definition
of $W_p$ in Lemma \ref{KF} we conclude that $W_p$ is $(m-1)$-dimensional and spacelike.
Thus $p$
lies on a smooth codimension-two, totally geodesic spacelike
surface of $\M$.
In these circumstances, the same construction performed by
R\'acz-Wald \cite{Racz-Wald}
in dimension four in order to find a canonical coordinate system near $p$
applies to arbitrary dimension. This gives a coordinate system $\{u,v,x^a\}$
($a,b=2,\cdots, m$) in an open connected
neighbourhood $U_p$ of $p$ with the following properties:
\begin{itemize}
\item The metric takes the form
\begin{eqnarray*}
\g = 2 G du dv + 2 v H_a dx^a du + g_{ab} dx^a dx^b
\end{eqnarray*}
with $G,H_a, g_{ab}$ smooth functions of $(uv,x^a)$, $G >0$
and $g_{ab}$ positive definite. 
\item The surface  $S_{p} \cap U_p$ takes the local
form $\{ u=0,v=0\}$.
\item The Killing vector $\xi$ reads $\xi = u \partial_u - v \partial_v$.
\item $\partial_v$ is future directed everywhere.
\end{itemize} 
Since $\partial_v$ is null and non-zero, the spacelike hypersurface
$\Sigma \cap U_p$  can be written as a graph $\{ v = \phi(u,x^a) \}$
(in particular, $\{u,x^a\}$ defines a local coordinate system on
$\Sigma \cap U_p$). Since $\Sigma$ is spacelike $\phi$ satisfies
$\partial_u \phi >0$ everywhere. 
Let $u_0 (x^a)$ be the solution of $\phi(u,x^a)=0$ (which exits because
$\phi$ vanishes on $p$).

Now, since $\lambda = 2\hat{G} u \phi$ where
$\hat{G} = G(u\phi,x^a)$ it follows
that either

\vs
 (i) $\Sigma^T \cap U_p = \{ u > 0 \} \cap \{ u > u_0\}$, or

\vs
(ii) $\Sigma^T \cap U_p = \{ u < 0 \} \cap \{ u < u_0\}$, or

\vs
(ii) $\Sigma^T  \cap U_p = ( \{ u > 0 \} \cap \{ u > u_0\} ) \cup
( \{ u < 0 \} \cap \{ u < u_0\} )$. 

\vs
\n  The corresponding Cauchy boundaries 
are:

\vs
For (i): $\{ u = \mbox{max} ( 0,u_0(x^a) ), x^a \} $

\vs
For (ii): $\{ u = \mbox{min} ( 0,u_0(x^a) ), x^a \}$

\vs
For (iii): The disjoint union of both.
\vs 

It is now obvious that the Cauchy boundary is a $C^0$ manifold (actually locally Lipschitz) without
boundary The uniqueness statement follows from hypothesis  {\small\bf A2}. \ep

 \section{Acknowledgments}

M.M. is very grateful to  Miguel S\'anchez for valuable
discussion on this subject.
The authors wish to thank the
Albert Einstein Institute and 
the University of Salamanca for their hospitality and support.
M.M. wishes the acknowledge financial support under the
projects FIS2009-07238 (Spanish MICINN)
and P09-FQM-4496 (Junta de Andaluc\'{\i}a and FEDER funds).

\section{Appendix A: Fixed points have $\Ione < 0$}
\label{I1}

In this Appendix we prove Lemma \ref{fixed}.   Let $p$
be a fixed point in $\tbd \Sigma^T$. In fact, the lemma
also holds for the more general case of $p \in \tbd \{\lambda > 0\}$.
For the purpose of the proof it is convenient to 
assume $p \in \tbd \{\lambda > 0\}$ and extend the definition of
horizon given in Definition \ref{horizon}
to any leaf ${\cal L}_{\alpha}$ intersecting $\tbd \{\lambda >0\}$. The constancy
of the surface gravity and Proposition \ref{surface} also hold for such horizons.

We know that $\Ione \leq 0$ on
$\{\lambda >0\}$ (see (\ref{expIone})) and, by continuity $\Ione(p) \leq 0$.
So, we only need to exclude the possibility $\Ione=0$. Let us 
assume that $\Ione |_p =0$ and find a contradiction.

\vs 
Our aim is to show that there exists a non-degenerate
horizon $\H_{\alpha}$ satisfying $p \in \overline{\H_{\alpha}}$. Since
$\Ione = - 2 \kappa_{\alpha}^2$ (see Proposition \ref{surface})
and $\kappa_{\alpha}$ is constant and non-zero
on a non-degenerate horizon, we 
would contradict $\Ione |_p =0$. 
To that aim we only need to find a
smooth path $\gamma (s)$ lying on a non-degenerate horizon
and containing $p$ in its closure.

\vs 
First we note that $d\Y|_{p} \neq 0$
and $dN|_p \neq 0$. Indeed, if $dN|_p=0$, then $Z|_p=0$ and the
definition of $\Ione$ (together with $\Ione=0$)
implies $d\Y|_{p}=0$. However, in Lemma 1 in
\cite{CarrascoMars2008} it is proved that a fixed point cannot have
$d\Y=0$ and $dN=0$ unless the Killing data $N,Y$ vanishes identically,
which is not the case (the proof of Lemma 1 in \cite{CarrascoMars2008} is done explicitly
in dimension $m=3$ but it carries through to arbitrary dimension with trivial
changes).
Thus, $dN|_p \neq 0$, and then the vanishing of $\Ione |_p$ also 
implies $d \Y|_p \neq 0$.

Now, in Lemma 8 in \cite{CarrascoMars2008} it is proved (again the proof
is done there in dimension 3, but extends to arbitrary dimension) 
that there exists 
a positive constant $b$ and a unit one-form $X \in T^{\star}_p \Sigma$
orthogonal to $dN|_p$ such that 
\begin{eqnarray}
d\Y |_p = \frac{2b}{Q} \left ( dN|_p \wedge X \right ),
\label{fij}
\end{eqnarray}
where $Q = |dN |_p |_{g}$. Evaluating $\Ione$ at $p$ we find
\begin{eqnarray}
\Ione |_p = 2 (b^2 - Q^2).
\label{IbQ}
\end{eqnarray}
Imposing $\Ione |_p=0$ we conclude $b = Q$. Let us now
evaluate the Hessian of $\lambda$ at $p$. Since
$\lambda = N^2 - |Y|^2_{g}$  and $p$ is a fixed point a
simple calculation 
yields
\begin{eqnarray*}
\Hess_g \lambda \, |_p = - 2 Q^2  X \otimes X.
\end{eqnarray*}
By the Gromoll-Meyer splitting Lemma \cite{Gromoll-Meyer}
there exist coordinates 
$\{y,x,z^A \}$ $A=3,\cdots, m$, in an open neighbourhood $W_p$ of $p$
such that 
$\lambda$ takes the form $\lambda = - Q^2 y^2 + f (x,z^A)$ on $W_p$
for some function $f$ and, moreover,
$p$ has coordinates $(0,\cdots 0)$, $X = dy |_p$,
$dN|_p =  dx |_p$ and  $f$ vanishes at $p$ together with its gradient and its
Hessian. From (\ref{fij})
we also have $\Y = Q (x dy - y dx ) + O(2)$.  We are now in a position
where the path $\gamma(s)$ mentioned above can be constructed.

For that we need
to investigate the region $\{ \lambda >0 \}$ near $p$. This region corresponds
to $f > Q^2 y^2$. Since $p \in \tbd \{ \lambda >0 \}$ 
it is clear that $p \in \overline{\{ f > 0 \} \cap \{y=0\}}$. If there
exists a smooth curve $\tilde{\Gamma}(s) \subset \{ y = 0 \}$ 
approaching $p$ and satisfying $f(\tilde{\Gamma}(s)) > 0$ then
we are done because the curve ${\Gamma}(s) : = \{ 
y= Q^{-1} \sqrt{f(\tilde{\Gamma}(s))}, f(\tilde{\Gamma}(s)) \}$ 
has the desired properties because (i) it lies on $\tbd \{\lambda > 0 \}$
and (ii) $d \lambda$ is nowhere
zero on the curve (because $y \neq 0$ there) and hence  $\Gamma(s)$
lies on a non-degenerate horizon.

So, it only remains to show that the curve $\tilde{\Gamma}(s)$ exists.
Define ${\cal V} \subset \{ y =0 \}$ as the set of points where the component
$\Y_y$ does not vanish. Since $\Y_y = Qx + O(2)$ it is clear that 
${\cal V}$ intersects $\{ f > 0 \}$ and also that there exists a smooth curve
$\tilde{\Gamma} (s)$ fully contained in ${\cal V}$ which approaches $p$.
Our last step is to show that in fact ${\cal V} \subset \{ f > 0 \}$. 
Consider a smooth curve $\hat{\Gamma}(s)$ starting on 
a point $q \in {\cal V} \cap \{ f > 0\}$. As long as $f$ remains
positive on this curve, consider the smooth curve 
$\overline{\Gamma} (s) \defi \{ y = Q^{-1}
\sqrt{f(\hat{\Gamma}(s))}, \hat{\Gamma}(s) \}$ which lies on a 
non-degenerate horizon. The $y$ component of
the equation $d \lambda = 2 \kappa_{\alpha} \Y$ on $\overline{\Gamma}(s)$
reads
\begin{eqnarray*}
- Q \sqrt{f(\hat{\Gamma}(s))} = \kappa_{\alpha} \Y_y
\end{eqnarray*}
Since $\kappa_{\alpha}$ is constant and $\Y_y \neq 0$ on ${\cal V}$
if follows that $f(\hat{\Gamma}(s))$ cannot become zero while remaining
inside ${\cal V}$. This implies ${\cal V} \subset \{ f >0 \}$ as claimed,
and the lemma is proved. \ep

\vs

\bibliography{biblio}
\bibliographystyle{plain}

\end{document}